\documentclass[letterpaper,dvipsnames]{article}
\pdfoutput=1
\usepackage{jheppub}
\usepackage{amsmath}
\usepackage{graphicx}

\usepackage{array}
\usepackage{ulem}
\usepackage{cancel}
\usepackage{xcolor}

\usepackage{slashed}
\usepackage{afterpage}
\usepackage{appendix}
\usepackage{multirow}
\usepackage{float}
\usepackage{mathtools}
\mathtoolsset{centercolon}

\usepackage{multirow}

\newcommand{\EM}{{\rm em}}

\newcommand{\tr}{\text{tr}}

\title{Testing the Heterotic String with the Axion-Photon Coupling}

\author[a,b]{Mario Reig}

\affiliation[a]{Theoretical Physics Department, CERN, 1211 Geneva 23, Switzerland}
\affiliation[b]{Rudolf Peierls Centre for Theoretical Physics, 
University of Oxford, Parks Road, Oxford OX1 3PU, United Kingdom}

\emailAdd{mario.reig.lopez@cern.ch}

\author[c,d]{and Timo Weigand}

\affiliation[c]{II. Institut f\"ur Theoretische Physik, Universit\"at Hamburg, Notkestrasse 9, 22607 Hamburg, Germany}

\affiliation[d]{Zentrum f\"ur Mathematische Physik, Universit\"at Hamburg, Bundesstrasse 55, 20146 Hamburg, Germany}

\emailAdd{timo.weigand@desy.de}

\abstract{ The discovery of an axion-like particle above the QCD line would rule out 
Grand Unified Theories, including the perturbative heterotic string with the Standard Model embedded in a single $E_8$ factor or $SO(32)$.  In this work we study a possible loophole to this observation, given by compactifications of the $E_8\times E_8$ heterotic string with a non-standard embedding of the Standard Model into the 10-dimensional gauge group.
If electromagnetism is embedded into both $E_8$ factors, axions can couple to photons via the anomaly without coupling to QCD.
We obtain upper bounds to the coupling-to-mass ratio $g_{a\gamma}/m_a$ for these axion-like particles as a function of the supersymmetry breaking scale and the unified gauge coupling. To be compatible with the measured gauge couplings and the weak mixing angle $\sin^2\theta_w$ at low-energies, phenomenologically viable models with non-standard $U(1)_Y$ embedding require sizeable one-loop threshold corrections from string states and/or charged matter at intermediate energy scales. We study how these effects modify the tree-level upper bounds to $g_{a\gamma}/m_a$ and show that, in the perturbative regime, they reduce the leading order estimates. Axion-like particles far above the QCD line are only possible in certain models where perturbation theory is lost.
The main conclusion is that the discovery of an axion violating the bounds found in this work would be incompatible with large classes of otherwise phenomenologically viable string models, including the perturbative heterotic  $SO(32)$ and $E_8\times E_8$ string, the type-I string, and certain heterotic M-theories.
 The role of small gauge instantons and worldsheet instantons in making some of the axion-like particles heavy and cosmologically relevant is briefly discussed.}

\begin{document}

\maketitle

\newpage

\section{Introduction and Summary}\label{sec:INTRO}
Originally conceived with the aim of describing the strong interaction, string theory has become the currently best understood framework where particle physics and gravity are unified in a quantum theory~(see \cite{Marchesano:2024gul} for a recent review). In a similar manner it was recognised that axions, originally proposed as a solution to the strong CP problem~\cite{Peccei:1977hh,Wilczek:1977pj,Weinberg:1977ma}, are generic predictions in string theory~\cite{Witten:1984dg,Choi:1985bz,Choi:1985je,Dine:1986bg,Svrcek:2006yi,Conlon:2006tq,Arvanitaki:2009fg}.\footnote{There are even arguments \cite{Dvali:2022fdv} that axions are a prediction of quantum gravity  more generally.} The reason is that the theory contains higher-form gauge fields that provide Goldstone bosons in the 4d effective field theory (EFT) when integrated over extra dimensions. In many instances, the predicted axions are perturbatively massless, only receiving a calculable potential from the explicit breaking of their global higher-form symmetries by the presence of non-perturbative extended objects in the spectrum~\cite{Svrcek:2006yi,Reece:2024wrn}. For this reason, string theory naturally provides an attractive solution to the so-called Peccei-Quinn \textit{quality problem}~\cite{Georgi:1981pu,Kamionkowski:1992mf,Barr:1992qq}~--~that is, the question: Why is the QCD axion light enough to solve the strong CP problem? 

In recent years, a large amount of effort both in theory and phenomenology has been dedicated to axions from string theory. This includes axions in the different string models including type IIA~\cite{Honecker:2013mya,Petrossian-Byrne:2025mto}, type IIB~\cite{Cicoli:2012sz,Cicoli:2013ana,Demirtas:2018akl,Hebecker:2018yxs,Mehta:2021pwf,Cicoli:2021gss,Carta:2021uwv,Cicoli:2021tzt,Demirtas:2021gsq,Cicoli:2022fzy,Gendler:2023kjt,Dimastrogiovanni:2023juq,Sheridan:2024vtt,Benabou:2025kgx,Yin:2025amn,Cheng:2025ggf}, the heterotic string~\cite{Choi:2011xt,Choi:2014uaa,Buchbinder:2014qca,Agrawal:2024ejr,Loladze:2025uvf,Leedom:2025mlr} as well as M-theory~\cite{Svrcek:2006yi,Im:2019cnl,DeLuca:2021pej}, and F-theory~\cite{Halverson:2019cmy,Halverson:2019kna}. With the goal of using  axion predictions as a tool to test string compactifications in cosmology, astrophysical systems and in the laboratory, this line of research resembles in some sense the spirit of the Swampland Programme~\cite{Vafa:2005ui, Brennan:2017rbf,Palti:2019pca,vanBeest:2021lhn}, which aims to find the theoretical constraints that a consistent EFT coupled to quantum gravity must satisfy. 

Identifying smoking gun observations or experiments which  rule out entire classes of quantum gravity theories continues to be very challenging. In this paper, extending the work of \cite{Agrawal:2022lsp, Agrawal:2024ejr}, we will argue that observing a light axion-like particle (ALP) whose coupling-to-mass ratio lies far above the QCD axion line in the ($g_{a\gamma}, m_a$) plane would 
 essentially rule out the {\it perturbative} heterotic string theory. Importantly, this result does not depend on the ALP being the cosmological dark matter (DM).

Previously, it had already been pointed out in \cite{Agrawal:2022lsp} that in Grand Unified Theories (GUTs)
in which the Standard Model (SM) gauge group is embedded into a single simple non-abelian Yang-Mills sector at high energies, ALPs coupled to photons necessarily lie at or below the QCD line. The reason is that in such constructions, only the QCD axion couples to photons via the anomaly and any other ALP can only receive an axion-photon coupling via mixing with the QCD axion.\footnote{Note that any additional contribution to the QCD axion potential, $V(a_{\rm QCD})=-\Lambda_{\rm QCD}^4\cos(a_{\rm QCD})-\Lambda_{\rm new}^4\cos(a_{\rm QCD}+\delta_{\rm CPV})$,  results in additional mass contributions to the QCD axion and spoils the solution to the strong CP problem unless the CP violating phase $\delta_{\rm CPV}$ is small enough. Even in cases where both $\Lambda_{\rm new}$ and $\delta_{\rm CPV}$ are tuned so that a cancellation in the axion mass is achieved, a light axion coupled to gluons well above the QCD line is subject to very stringent constraints~\cite{Hook:2017psm,Balkin:2022qer,Kumamoto:2024wjd,Gomez-Banon:2024oux}.} 
Importantly, however, such mixing-induced couplings of light ALPs come with the associated mixing suppression, which for light axions scales as $\sim m_{\rm light}^2/m_{a_{\rm QCD}}^2$ (see also~\cite{Gavela:2023tzu,Dunsky:2025sgz}). Irrespective of its mass, it turns out that  the ALP coupling-to-mass ratio cannot exceed that of the QCD axion \cite{Agrawal:2022lsp}.
 This result carries over to perturbative heterotic string vacua in which the SM sector is embedded into a single $E_8$ (or SO(32)) factor at the compactification scale \cite{Agrawal:2024ejr}. This leaves as a potential loophole the possibility that one of the SM gauge group factors, for example hypercharge, receives a contribution from both $E_8$ factors of the heterotic string. 
 In this work, we analyse such constructions and conclude that an ALP far above the QCD line is possible only at the price of abandoning the regime of controlled perturbative string theory. In this sense, finding an ALP whose coupling-to-mass ratio, $g_{a\gamma}/m_a$, is much larger than that of the QCD axion rules out the realm of the perturbative heterotic string. 

\subsection{Review: Axions in perturbative heterotic string theory} \label{sec:reviewaxions}
The heterotic string~\cite{Gross:1984dd} is one of the most promising frameworks to obtain gauge theories that resemble the Standard Model (SM).
  Among its most striking features is the fact that the 10d gauge group is restricted to be $E_8\times E_8$ or $SO(32)$ by the requirement of anomaly cancellation~\cite{Green:1984sg,Adams:2010zy,Kim:2019vuc}.  The EFT of heterotic string theory is 10d $\mathcal{N}=1$ supergravity (SUGRA) coupled to the super Yang-Mills (YM) sector.  In addition to the dilaton (setting the string coupling) and the graviton, it contains a two-index antisymmetric tensor, $B_2$. We will focus on heterotic models compactified on smooth Calabi-Yau (CY) manifolds, but since our results rely on anomaly matching and, relatedly, how the different axion linear combinations couple to the different instantons in the spectrum, we expect these results to hold for orbifold theories and other constructions.

When $B_2$ and its 10d dual, $B_6$, are integrated over 2-cycles $C_i$ and the 6d compact space, $X$, 
\begin{equation}\label{eq:MI_and_MD_axions}
   a=\int_{X}B_6 \,,\,\,\,\,\, b_i=\int_{C_i}B_2\,,
\end{equation}
one obtains axions in the 4d EFT. These are the model-independent axion, $a$, and model dependent axions, $b_i$. The shift symmetries of these fields are protected by the global 6-form and 2-form symmetries of $B_6$ and $B_2$, respectively, only broken by NS5-branes~\cite{Strominger:1990et} (for simplicity we will refer to them simply as {5-branes}) and worldsheet instantons~\cite{Wen:1985jz}.  In a similar way to \textit{predicting} gravity, the EFT of the heterotic string unavoidably also contains axions and high-quality PQ-like symmetries.\footnote{The quality of these axions may be altered in the presence of worldsheet instantons with small action, instantons of other confining interactions, or even small QCD instantons if there exist multiple coloured states at high energies~\cite{Dine:1986bg}. We will quantify these effects in later sections.} The number of  these axions depends on the topology of the internal space, in particular on its second Betti number as $N_a=1+h_{11}$.

The requirement of anomaly cancellation determines the 10d gauge group and also uniquely fixes how $B_2$ couples to gauge bosons and gravitons via the celebrated Green-Schwarz counter-term~\cite{Green:1984sg}. In the conventions in~\cite{Svrcek:2006yi}, the latter reads
\begin{equation}\label{eq:GS}
  S_{\rm GS}= -\frac{1}{768\pi^3} \int B_2\wedge X_8\,.
\end{equation}

The concrete form of $X_8$ depends on the 10d gauge group and is fixed by anomaly cancellation.
For example, for the heterotic string with 10d gauge group $E_8\times E_8$, the eight-form in Eq.~\eqref{eq:GS} is given by~\cite{Polchinski:1998rr}
\begin{equation}
    X_8 = -(\tr _1\mathcal{F}^2+\tr _2\mathcal{F}^2)\tr \mathcal{R}^2+2\left[(\tr _1\mathcal{F}^2)^2+(\tr _2\mathcal{F}^2)^2-\tr _1\mathcal{F}^2\tr _2\mathcal{F}^2\right]+\tr \mathcal{R}^4 +\frac{1}{4}(\tr \mathcal{R}^2)^2\,.
\end{equation}
Here $\mathcal{F}$ is the 10d gauge field strength, with $\tr_i$ being the trace in the fundamental representation of $E_8^{(i)}$, and $\mathcal{R}$ is the Riemann curvature 2-form. 
 This implies that if the heterotic string is the UV completion of the SM, axions must couple to gauge bosons and a candidate for the QCD axion is naturally predicted. Furthermore, the low-energy axion-photon coupling is uniquely determined by the embedding of the SM gauge group into the 10d gauge group, be it $E_8\times E_8$ or $SO(32)$~\cite{Agrawal:2024ejr}. The topological nature of the axion-gauge boson coupling  makes these predictions robust against any physics at intermediate scales or different symmetry breaking patterns. Axions therefore provide an opportunity to test the heterotic string in cosmology, astrophysics, and in the lab.

In addition to model independent and model dependent axions, in scenarios with pseudo-anomalous $U(1)$ gauge symmetries~\cite{Dine:1987xk}, it is also possible to have axions from the phases of complex scalar fields, $\Phi_j=\hat{\phi}_j e^{iq_j c_j}$, after they develop a non-zero vacuum expectation value (VEV)~\cite{Svrcek:2006yi,Choi:2011xt,Cicoli:2013ana,Cicoli:2013cha,Petrossian-Byrne:2025mto,Loladze:2025uvf}. Their phases, $c_j$, mix with the model independent and model dependent axions when these shift under the anomalous $U(1)$ symmetries,
\begin{align}\label{eq:different_kinds_of_axions}
   &a  \rightarrow a  + \sum_n \lambda^{(n)}  q_{a}^{(n)} \, ,\,\,\,\,\,\,
   b_i(x)\rightarrow b_i + \sum_n\lambda^{(n)}q_{i}^{(n)} \, ,\,\,\,\,\,\,
   c_j \rightarrow c_j+ \sum_n \lambda^{(n)}  q_{c_j}^{(n)} \, .
\end{align}
Here $n$ indexes the $U(1)$ factors in the effective theory, the $q$'s are the $U(1)$ charges of each axion, and the $\lambda^{(n)}$ are the gauge transformation parameters.  
A linear combination of $a$, $b_i$ and $c_j$ is eaten by the anomalous gauge boson and the orthogonal remains perturbatively massless. As shown in \cite{Agrawal:2024ejr}, the number of (light) axions does not change by having anomalous $U(1)$ symmetries -- for every anomalous $U(1)$ there will be a combination of phases $c_j$, $a$, and $b_i$ that compensates for the eaten axion. 
The coupling to gauge bosons of the uneaten linear combination  is inherited from the mixing of $c_j$ with $a$ and $b_i$ via the GS term~\eqref{eq:GS}.

Finally, another source of axions that has received less attention in the  literature are the spacetime-filling heterotic 5-branes~\cite{Strominger:1990et} which may be needed in order to satisfy the Bianchi identity, see Eq.~\eqref{Biachni-Id-gen}. These objects are non-perturbative in nature and contain additional 2-form fields (different from the 10d $B_2$) living in their worldvolumes which we call $\tilde B_2^{(r)}$. These fields also provide axions in the 4d EFT when integrated over 2-cycles,
\begin{equation}
   \tilde b_r = \int_{\Gamma_r} \tilde B_2^{(r)} \,. 
\end{equation}
For the sake of completeness, in  Appendix~\ref{App:non_perturbative_axions} we derive the couplings of these non-perturbative axions $\tilde b_r$ to gauge bosons, following~\cite{Blumenhagen:2006ux}.
 
\subsection{ALP coupling-to-mass ratios in heterotic GUTs}

Let us now consider a simple situation where 
 the gauge background embedded into the original gauge group does not
include non-trivial $U(1)$ gauge fluxes.\footnote{Turning on non-trivial $U(1)$ fluxes within the 10d gauge group, as studied systematically in \cite{Blumenhagen:2005ga,Blumenhagen:2006ux}, can result in more complex scenarios and will be the topic of this work.} Below the compactification scale, and once the gauge bosons associated to generators broken by the gauge bundle and  discrete Wilson lines have been integrated out, it can be shown~\cite{Agrawal:2024ejr} that the most general axion EFT consists of two different linear combinations of axions coupling as
\begin{equation}\label{eq:4dEFT_axion_couplings}
   \mathcal{L=}  \frac{\theta_1}{8\pi^2} \tr_1 F^2 +  \frac{\theta_2}{8\pi^2} \tr_2 F^2 \,.
\end{equation}
Here $\tr_{1,2} F^2=\sum_jk^{(1,2)}_j\tr \left ( F^{(1,2)}_j\wedge F^{(1,2)}_j\right )$ include the unbroken 4d gauge group factors coming from the first and second $E_8$, respectively. The coefficients $k_i^{(1,2)}$ are the levels of embedding of the $i-$th gauge group into $E_8^{(1,2)}$. The linear combination that couples to each set of gauge bosons, $\tr_{1,2} F^2$, is given by
\begin{equation}\label{eq:linear_comb_theta_12}
    \theta_{1,2}= a + \sum_i n_i^{(1,2)} b_i+ \sum_j 
 m_j^{(1,2)}c_j\pm \sum_r N_r\tilde b_r\,,
\end{equation}
with $n_i,m_j$ being calculable anomaly coefficients that depend on the properties of the manifold and $N_r$ is the number of 5-branes wrapping the two-cycle, $\Gamma_r$.\footnote{Note that in the absence of NS5-branes, $N_r=0$ and one can use the Bianchi identity to obtain $n_{i}^{(1)}=-n_{i}^{(2)}=n_i$. See Appendix~\ref{App:non_perturbative_axions} for the most general case.} We see that in general $\theta_{1,2}$ are complicated linear combinations of the model independent axion $a$, model dependent axions $b_i$, field theory axions $c_j$, and non-perturbative axions from 5-branes, $\tilde b_r$. While the shape of the concrete linear combination depends on the compactification, the fact that there exist only two linear combinations coupled to gauge bosons, $\theta_{1,2}$, is a model independent prediction of the $E_8\times E_8$ heterotic string with GUT embeddings into the same $E_8$.

This has important consequences for phenomenology: If the SM is embedded into the same $E_8$ factor, then any axion coupled to photons necessarily satisfies the bound~\cite{Agrawal:2024ejr}
\begin{equation}\label{eq:coupling_mass_relation}
    \frac{g_{a\gamma}}{m_a}\lesssim \frac{\alpha_{\rm em}}{2\pi}\left (\frac{E}{N}-1.92\right )\frac{1}{f_\pi m_\pi}\,,
\end{equation}
with the inequality being saturated for the QCD axion. The ratio of anomaly coefficients is given by $E/N=(k_1+k_2)/k_3$, with $k_i$ the level of embedding of the $i-$th SM gauge group into the simple group. In the heterotic string this ratio is not necessarily restricted to the standard 4d GUT prediction, $E/N=8/3$~\cite{Srednicki:1985xd}, and can take values in the range $2\lesssim E/N \lesssim 63/2$.  
The inequality~\eqref{eq:coupling_mass_relation} has important experimental implications: Finding an axion above the QCD line (that is, violating the inequality  (\ref{eq:coupling_mass_relation})), is incompatible with heterotic string theory where  the SM is embedded into the same $E_8$. The same result also applies to heterotic $SO(32)$ as well as type-I string theory whenever the SM is embedded into the 10d $SO(32)$ sector. Analogous constraints for $g_{a\gamma}/m_a$ have been obtained for 4-dimensional Grand Unified Theories~\cite{Agrawal:2022lsp} as well as in orbifold GUT theories~\cite{Agrawal:2025NEW}. 

In the heterotic string the axion decay constant for the model independent and model dependent axions are predicted to be close to the GUT scale~\cite{Svrcek:2006yi} unless field theoretic axions are introduced. The latter allow for low-scale decay constants~\cite{Buchbinder:2014qca} as well as the formation of a network of cosmic axion strings~\cite{Loladze:2025uvf,Petrossian-Byrne:2025mto} if the reheating temperature is high enough. This is the so-called \textit{post-inflation} axion scenario -- a  situation where the QCD axion mass can be predicted from string-network simulations which currently point towards $O(10-100)\,\mu$eV \cite{Saikawa:2024bta,Benabou:2024msj,Correia:2024cpk}. However, in heterotic string theory this is only expected to occur in certain regions of moduli space. Therefore, in minimal realizations of the heterotic string we expect the QCD axion to have a decay constant around the GUT scale corresponding to a mass around $m_a\sim 10^{-9}$ eV  (see~\cite{Leedom:2025mlr} for a recent, comprehensive study of masses and effective couplings). For this reason heterotic axions with GUT scale decay constant constitute a well-motivated target for experiments including the ABRACADABRA~\cite{Kahn:2016aff,Ouellet:2018beu,Ouellet:2019tlz,Salemi:2021gck} and DM-Radio experiments~\cite{DMRadio:2022jfv,DMRadio:2022pkf,Benabou:2022qpv,DMRadio:2023igr}, superconducting cavities~\cite{Berlin:2019ahk,Giaccone:2022pke} and CASPER~\cite{Graham:2013gfa,Budker:2013hfa,JacksonKimball:2017elr,Aybas:2021cdk,Dror:2022xpi}. See also~\cite{Benabou:2025kgx} for a recent discussion in the context of type IIB strings and other higher-dimensional axion models.

\subsection{Summary of results}

The discussion so far has left an interesting potential loophole to the conclusion that ALPs in heterotic models are subject to the bound (\ref{eq:coupling_mass_relation}): 
 The SM gauge factors may receive contributions from the two $E_8$ factors. In particular this can be the case for hypercharge \cite{Blumenhagen:2006ux}.
  In this article we therefore consider classes of heterotic string models with a non-standard embedding of electromagnetism into both $E_8$ groups and study the impact on low-energy axion couplings. For definiteness, we will assume that $SU(3)_C$ and $SU(2)_w$ come from the first $E_8$, as in~\cite{Blumenhagen:2006ux}.
In this case, $\theta_1$ obtains an IR potential from its coupling to gluons and becomes the QCD axion.\footnote{We assume that $\theta_1$ obtains a mass mainly from QCD and other effects are subdominant. Nevertheless, our findings do not rely on the quality of $\theta_1$.} On the other hand, as the SM hypercharge $U(1)_Y$ receives contributions from $U(1)$ factors in both $E_8$'s, $\theta_2$ behaves as an ALP and couples to photons without coupling to gluons. 
 Such constructions, strictly speaking, leave the framework of a GUT and for this reason we expect features that do not appear in a framework in which the SM originates from one simple non-abelian gauge group. In particular, this implies the appearance of more than one ALP and affects the details of the ALP coupling-to-mass ratio.
 The question therefore becomes, what is the ALP potential and does it lead to a violation of the bound  (\ref{eq:coupling_mass_relation})? 
 Remarkably, we do find predictive constraints even in this considerably more general class of heterotic models.
 The point is that such models are still \textit{approximate} GUTs, at least within the perturbative regime, and hence the general mechanism of \cite{Agrawal:2022lsp,Agrawal:2024ejr} continues to be effective, with changes that we quantify.
 
\begin{figure}[t]
    \centering
    \includegraphics[scale=0.4]{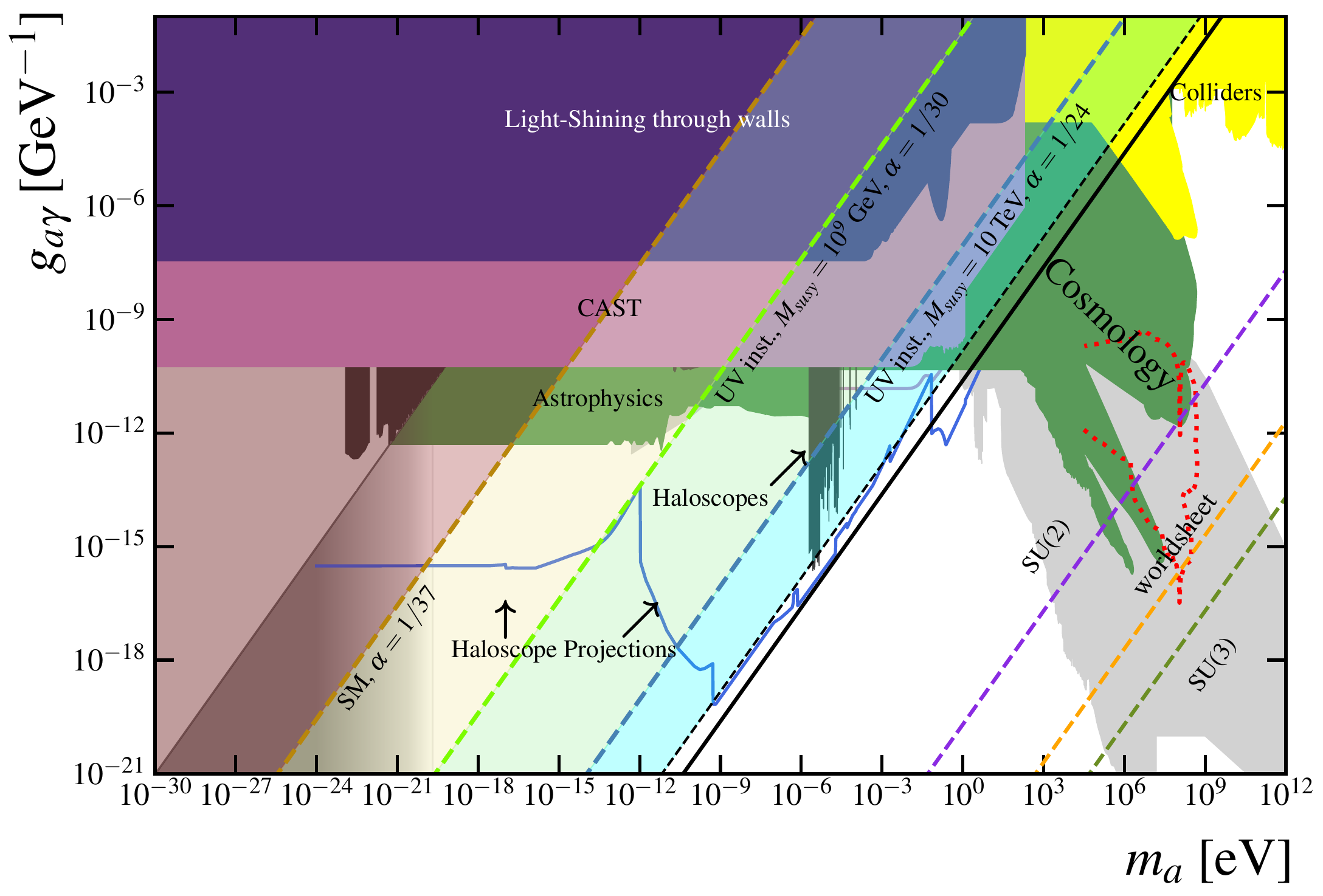}
    \caption{\textbf{Summary of results}. Axion parameter space for the heterotic string axiverse with non-standard embedding of hypercharge. Experimental contraints adapted from \cite{AxionLimits}. The black dashed line corresponds to the standard GUT QCD axion prediction with $E/N=8/3$. The black solid line corresponds to the QCD axion prediction with $E/N=49/24$, which exhibits an accidental cancellation with the pion mixing contribution that leads to $g_{a\gamma}$ a factor $\sim 6$ smaller than the standard GUT prediction. Different dashed lines correspond to different $g_{a\gamma}/m_a$ predictions for $\theta_2$ and $\varphi$. (As we are neglecting the effect of threshold corrections in the ALP mass, we have a comparable ratio for both ALPs, $g_{\theta_2\gamma}/m_{\theta_2}\sim g_{\varphi\gamma}/m_{\varphi}$.)  The \textcolor{Plum}{purple} and \textcolor{OliveGreen}{dark green} dashed lines correspond to unbroken $SU(2)$ and $SU(3)$ groups in the second $E_8$. 
    The \textcolor{cyan}{light blue} dashed line corresponds to the second $E_8$ fully broken to $U(1)$s and $\alpha_{\rm UV}=1/24$ and $M_{\rm susy}=10^4$ GeV.
    The \textcolor{green}{light green} dashed line corresponds to $\alpha_{\rm UV}=1/30$ and $M_{\rm susy}=10^9$ GeV.
    The \textcolor{Tan}{gold} dashed line corresponds to a situation where there is no low-scale supersymmetry and the gauge couplings are given by extrapolating the measured values at the weak scale assuming SM matter content, $\alpha_{\rm UV}=1/37$. 
In any of the three cases, the charged matter spectrum is given by 3-generation MSSM-like models.
    The shaded regions to the right of each of the $g_{a\gamma}/m_a$ lines can be populated, for example via axion mixing, in each situation, e.g. the shaded light green region is allowed for the cases $\alpha_{\rm UV}=1/30$ and $M_{\rm susy}=10^9$ GeV, and for $\alpha_{\rm UV}=1/37$ but not for $\alpha_{\rm UV}=1/24$ and $M_{\rm susy}=10^4$ GeV. The \textcolor{BrickRed}{shaded red} region is incompatible with the heterotic string even with non-standard $U(1)_Y$ embedding.
    \textcolor{orange}{Orange} corresponds to the prediction for a model-dependent axion receiving its mass from a worldsheet instanton with action $S_{ws}\sim 2\pi \text{Vol}(C)$ and 2-cycle volume $\text{Vol}(C)=10$. As shown later in figures~\ref{fig:ParameterSpace_with_small_k1} and \ref{fig:ParameterSpace_with_threshold_correction} any mechanism that allows to recover the measured gauge couplings (e.g. small level of hypercharge embedding or threshold corrections) tend to decrease the leading order estimates for $g_{a\gamma}/m_a$ in this figure.}
       \label{fig:ParameterSpace}
\end{figure}

To compute the irreducible\footnote{There could exist additional gauge instantons generating an axion potential on top of $V(\theta_2)$. These will generate additional contributions to the ALP mass as we will see later.} ALP potential we will consider a class of models where the second $E_8$ is completely broken by the embedding of the gauge bundle, 
\begin{equation}
    E_8^{(2)}\rightarrow \prod_m U(1)_m \,.
\end{equation}
Incidentally, in these models, the topologically non-trivial line bundles introduce a new kind of axion linear combination previously overlooked in the literature, $\varphi$, that only couples to the $U(1)$ gauge symmetries in the 4d EFT. We will see that in some cases this axion, which for reasons that will become clear later we call \textit{non-universal} ALP, can also couple to photons. 
Avoiding chiral exotics -- that is, chiral fermions with exotic SM charges -- offers non-trivial restrictions on these models. Even in absence of a confining gauge interaction from the second $E_8$, the potential $V(\theta_2,\varphi)$ is not flat. Shift-symmetry breaking effects are induced by Euclidean NS5-branes~\cite{Strominger:1990et} or, equivalently, by UV (small) gauge instantons at the compactification scale that are calculable. In some cases, these effects can be reliably estimated just from low-energy observables such as gauge couplings.
As a result, in  Figure~\ref{fig:ParameterSpace}, we obtain an upper bound to the coupling-to-mass ratio $g_{a\gamma}/m_a$ at \textit{tree-level} for any axion coupled to photons. 
 Note that, crucially, the ratio $g_{a\gamma}/m_a$ does not depend on the axion decay constant and only requires knowledge of the different non-perturbative objects that break the shift-symmetry of the axion. These bounds apply to heterotic models with non-standard hypercharge embedding, type-I string theory as well as the strong coupling limit of $E_8\times E_8$ string theory as long as the SM originates from the perturbative part of the gauge sector.

In their minimal realisation, the models with a non-standard embedding of hypercharge that we consider are known to predict an incorrect tree-level value for the weak mixing angle, $\sin^2\theta_w$, and the UV gauge couplings \cite{Blumenhagen:2006ux}. As a prototype example we study an explicit model where $\alpha_2=\alpha_3=\alpha_{\rm GUT}=\frac{8}{3}\alpha_Y$ at the GUT scale. In theories where the only light matter consists of three chiral SM families this prediction is a phenomenological disaster. Interestingly, there exist a few ways out to solve this issue and to recover the measured gauge couplings at low-energies. These include one-loop threshold corrections from heavy string states -- which are the actual reason for the existence of the \textit{non-universal} ALP
--, new states with SM charges between the weak and the KK scale, as well as a small level of embedding of $U(1)_Y$ into the second $E_8$. In Section~\ref{sec_thresholds} we argue that, generically, the value of the upper bound to $g_{a\gamma}/m_a$ is further reduced compared to the leading order analysis of Figure~\ref{fig:ParameterSpace} as we recover the standard GUT prediction for the gauge couplings and $\sin^2\theta_w$. These results are shown in Figures \ref{fig:ParameterSpace_with_small_k1} and \ref{fig:ParameterSpace_with_threshold_correction} for two prototypical examples. As we discuss, the only way to violate this conclusion is by leaving the regime of perturbative control, in which, however, the EFT analysis breaks down and a new assessment in a duality frame different from the perturbative heterotic string is required.

Our results offer well-defined regions in the $(g_{a\gamma}, m_a)$  parameter space where axions \textit{can and cannot live} in perturbative heterotic strings with a non-standard embedding of the SM into both $E_8$'s. As one can see from the different lines for the coupling-to-mass ratio in Figures \ref{fig:ParameterSpace}, \ref{fig:ParameterSpace_with_small_k1}, \ref{fig:ParameterSpace_with_threshold_correction},
the bounds become considerably stronger in the presence of additional new physics at low or intermediate scales which modifies the running of the gauge couplings compared to the SM.
These bounds to $g_{a\gamma}/m_a$ are of great importance given the large experimental programme trying to find the axion at current and future experiments~\cite{Marsh:2018dlj,PhysRevLett.118.091801,Lawson:2019brd,Beurthey:2020yuq,Schutte-Engel:2021bqm,DMRadio:2022pkf,Aja:2022csb,Bourhill:2022alm,ALPHA:2022rxj,Oshima:2023csb,DeMiguel:2023nmz,Ahyoune:2023gfw,Alesini:2023qed,BREAD:2023xhc,CAST:2024eil,Kalia:2024eml,Friel:2024shg,Koppell:2025dmt,Baryakhtar:2025jwh}. In the event of a discovery, our results demonstrate that an axion signal could 
rule out entire classes of string models. In particular we determine regions of parameter space where finding an axion would be incompatible with perturbative heterotic string theory -- independently of the embedding of the SM group --, type-I string theory, and also some strongly coupled heterotic M-theories. Such a discovery would therefore leave non-unified type-II string models, F-theory models, and similar, possibly to date unknown, constructions without a GUT origin of the SM (see the recent review \cite{Marchesano:2022qbx} and references therein for details of such constructions) as the only viable string UV completions of the SM. In addition, we also find that heavy axions coupled to photons with $g_{a\gamma}/m_a$ much smaller than the QCD axion are generically expected in the heterotic string axiverse with interesting implications for cosmology.

We acknowledge that the properties of axions, similar to most of the low-energy predictions of string compactifications, depend on the concrete mechanism for moduli stabilisation. However, the physics associated to stabilizing the moduli fields is not expected to alter the topological properties of axion couplings. While their masses could be affected (especially in supersymmetric mechanisms where axions gain the same mass as the real part of the complex moduli), these typically will only be enhanced with respect to the non-perturbative estimates that we provide. In this way, we expect that our predictions for $g_{a\gamma}/m_a$ can, at most, decrease once moduli stabilization is implemented.

This article is structured as follows. In Section \ref{sec:line_bundle_GUT_models} we review heterotic GUT models with non-standard hypercharge embedding and study the conditions to avoid chiral exotics. We also discuss how threshold corrections as well as new matter at intermediate scales may help to improve gauge coupling unification. 
 We identify two ALPs, $\theta_2$ and $\varphi$, coupling to electromagnetism as in (\ref{eq:eff_Lag_non-standard_embedding}).
 In Section \ref{sec:ALP_potential}, we compute the (leading order) irreducible axion potential and $g_{a\gamma}/m_a$ for any axion in heterotic string theory.
  In models with a remnant non-abelian confining gauge sector, $\theta_2$ receives a large mass. 
  Both $\theta_2$ and $\varphi$ receive masses from small UV instantons associated with Euclidean NS5-branes, whose suppression is controlled by the abelian gauge couplings.
  In addition we comment on worldsheet instantons which, in regions of moduli space where threshold corrections are small, give sizeable contributions to the axion masses.
   In Figure \ref{fig:ParameterSpace} we summarise the coupling-to-mass ratio
for ALPs taking into account the tree-level values of the gauge couplings. 
   Section~\ref{sec_thresholds} is dedicated to showing how mechanisms that allow to obtain the correct gauge couplings and $\sin^2\theta_w$ tend to decrease the tree-level prediction for the ratio $g_{a\gamma}/m_a$. The only exception to this general statement is by allowing large threshold corrections of a certain sign choice. This is shown in Figures \ref{fig:ParameterSpace_with_small_k1} and \ref{fig:ParameterSpace_with_threshold_correction}. The conclusion is that a significant violation of the bound (\ref{eq:coupling_mass_relation}) requires thresholds so large that the model lies outside the perturbative regime.
 Finally, we discuss the implications for phenomenology and conclude in Section \ref{sec:conclusions}. 
 The discussion of concrete models and of the 5-brane axion $\tilde b_r$ is relegated to the appendices.

\section{Heterotic GUT models with non-standard hypercharge embedding}\label{sec:line_bundle_GUT_models}
In this section we study a class of $E_8 \times E_8$ heterotic models in which hypercharge is embedded  into both $E_8$ factors in a non-standard way \cite{Blumenhagen:2006ux}. $SU(3)_C$ and $SU(2)_w$ are embedded at level one into the first $E_8$ and the SM fermions descend from gauginos in this group. As we will discuss, for the ALP $\theta_2$ defined in (\ref{eq:linear_comb_theta_12}) to remain as a light axion coupled to photons there should not arise any non-abelian sector from the second $E_8$. Furthermore, consistency with basic phenomenological requirements forbids chiral fermions from gauginos in the second $E_8$ with non-zero hypercharge. We will see that both features can be achieved in models 
 of the type introduced in \cite{Blumenhagen:2006ux} that satisfy a set of non-trivial conditions. We will derive these conditions and provide a concrete example. 

{A general property of  heterotic models with non-standard hypercharge embedding on smooth Calabi-Yau threefolds is that they  predict incorrect tree-level values for the gauge couplings in the UV. Fortunately, in these models threshold corrections from heavy string modes are different for the abelian and non-abelian 4d gauge groups \cite{Blumenhagen:2005ga}. We will review how these can help with gauge coupling unification. Separately we will also show explicitly how a new kind of ALP appears via these non-universal threshold corrections. To simplify the discussion, the study of this new kind of ALP and its couplings will be postponed to section~\ref{sec_varphi}.}

\subsection{Realistic models with non-standard hypercharge embedding}
\label{sec:Timo's_model}
A class of $E_8 \times E_8$ heterotic Calabi-Yau compactifications where the SM hypercharge is embedded in both $E_8$ factors was  introduced in~\cite{Blumenhagen:2006ux}. These constructions can give rise to the SM gauge group and a realistic fermion spectrum even on a smooth Calabi-Yau threefold $X$ which is simply connected and hence does not admit non-trivial Wilson lines. The general idea is to break both $E_8$ factors to a product of non-abelian and abelian gauge groups,
  \begin{equation}\label{E81E82-gen}
  E_8^{(1)} \times E_8^{(2)}  \to (G_1 \times \prod_{m_1} U(1)_{m_1} )   \times (G_2 \times \prod_{m_2} U(1)_{m_2} ), 
  \end{equation}
 where the non-abelian parts $G_1$ and $G_2$ can themselves factorise further. The gauge background on the smooth Calabi-Yau threefold $X$ inducing such a breaking can be identified as the field strength of two vector bundles $W_1$ and $W_2$  with $c_1(W_i)=0$ which are in general direct sums of non-abelian gauge bundles and line bundles. Embedding their structure groups into $E_8^{(1)}$ and $E_8^{(2)}$ breaks the four-dimensional gauge group to the commutant within $E_8 \times E_8$. 
 As usual, the gauge background is subject to the Bianchi identity 
\begin{equation} \label{Biachni-Id-gen}
    {\rm tr}_{1} {\bar F}^2 + {\rm tr}_{2} {\bar F}^2 - {\rm tr} {\bar R}^2 = [W] \,,
\end{equation}
where $[W]$ is an effective curve class on the compactification space $X$ that is wrapped by spacetime-filling heterotic 5-branes, and $\bar F$ and $\bar R$ denote the components of the field strength and curvature 2-form along $X$, respectively. Furthermore, since both $W_i$ are direct sums of vector and line bundles with non-zero first Chern classes (such that $c_1(W_i) =0$), a D-term supersymmetry condition arises \cite{Blumenhagen:2005ga, Blumenhagen:2006ux} that is required for stability at the compactification scale.

 In the context of realistic model building, the non-abelian group $G_1$ appearing in (\ref{E81E82-gen}) can either be a GUT group or directly contain the SM non-abelian groups $SU(3)_C \times SU(2)_w$. 
The $U(1)$s  from both $E_8$ factors are in general massive via a St\"uckelberg mechanism, but certain linear combinations can remain massless. This way, hypercharge can be identified as a linear combination of $U(1)$ factors from both $E_8$.
 
 We denote by $L_{m_i}$ a line bundle embedded into $E_8^{(i)}$. Since its abelian structure group commutes with itself, the visible gauge sector contains a corresponding factor $U(1)_{m_i}$. 
 Its 4d  field strength $f_{m_i}$ interacts with the 2-forms $b_I$ dual to the axions $a_I$ via a St\"uckelberg-type coupling of the form
   \begin{equation}
   S_{\rm St.} = \int_{\mathbb R^{1,3}}  \sum_{I, m_i} M_{I m_i}    f_{m_i} \wedge b_I  \,.
   \end{equation}
Here $I= 0, i, r$ labels the model independent axion $a$, the model dependent axions $b_i$ and the axions $\tilde b_r$ obtained from the spacetime filling NS5-branes needed to satisfy the Bianchi identity, see Section \ref{sec:reviewaxions}.
 This coupling induces a St\"uckelberg mass for the abelian gauge bosons, with a mass matrix 
 \begin{equation} \label{massmatrix-1}
{\cal M}_{m_i,m_j} \sim  \sum_I M^T_{m_i I}   M_{I m_j} \,.
 \end{equation}
 The coupling matrix is given by 
\begin{eqnarray} \label{massmatrix}
M_{I m_i} = \kappa_{m_i,m_i} \int_X c_1(L_{m_i}) \wedge \omega_I \,,
\end{eqnarray}
 where the precise form of $\omega_I \in H^4(X)$  can be found in \cite{Blumenhagen:2006ux} 
 but plays no role for us.
 Crucial for our analysis, however, are the group theoretic invariants $\kappa_{m_i,m_i}$ appearing in (\ref{massmatrix}), and  related invariants
 $\eta_{m_i,n_i}$ which we will encounter later. Both are defined as
 \begin{equation}   \label{traces-inv.}
{\rm Tr}_{E_8^{(i)}} {F}_i {\bar {F}}_i = \sum_{m_i,n_i} \kappa_{m_i,n_i} f_{m_i} \bar f_{n_i}  \,,  \qquad 
{\rm Tr}_{E_8^{(i)}} {F}_i^2 = 2 \,  {\rm Tr}_{G_i} {F}_i^2 + \eta_{m_i,n_i} f_{m_i} f_{n_i} \,.
 \end{equation}
Here ${F}_i$ and $\bar {F}_i$ denote the 4d and internal components of the $E_8^{(i)}$ field strengths while $f_{m_i}$ and $\bar f_{m_i}$ are the 4d and internal abelian field strengths, respectively. A linear combination 
 \begin{equation}
\sum_{m_i, i=1,2} \alpha_{m_i} U(1)_{m_i} \,,\, \text{ remains massless iff:} \quad M_{I m_i} \alpha_{m_i} = 0 \,.
\end{equation}
  
For example, to break $E^{(1)}_8$ directly to the SM gauge group\footnote{Other examples of phenomenologically interesting constructions studied in \cite{Blumenhagen:2006ux} include flipped SU(5) models, in which $E^{(1)}_8$ is broken to $SU(5) \times U(1)_0$ by an $SU(4) \times U(1)$ bundle.}
one can consider a gauge background corresponding to the direct sum  \cite{Blumenhagen:2006ux}
\begin{equation} \label{SMemb-1}
W_1 = V \oplus L_0^{-1} \,,
\end{equation}
of a $U(5)$ vector bundle $V$ and a line bundle $L_0^{-1}$ with  $c_1(V) = c_1(L_0)$  in such a way that it induces the breaking 
\begin{equation} \label{E81breaking}
E_8^{(1)} \to SU(3) \times SU(2) \times U(1)_0 \,.
\end{equation}
  Since the $U(1)_0$ gauge potential acquires a St\"uckelberg mass, another abelian gauge group factor is needed to obtain a massless linear combination that can be identified with hypercharge $U(1)_Y$.
 The simplest way to achieve this is by embedding 
  in the second $E_8$ the direct sum $W_2 = L_1 \oplus L_1^{-1}$
 such as to break  
 \begin{equation} \label{E82-E7}
 E_8^{(2)} \to E_7\times U(1)_1 \,.
\end{equation}
 If we take, for simplicity, 
 \begin{equation}
L_0 = L_1 \,,
 \end{equation}
 the mass matrix (\ref{massmatrix-1}) has rank one so that only one linear combination of $U(1)_0$ and $U(1)_1$ becomes massive by the St\"uckelberg mechanism \cite{Blumenhagen:2005ga, Blumenhagen:2006ux}. 
   The orthogonal combination given by
 \begin{equation} \label{lin-comb1}
U(1)_0 - \frac{\kappa_{0,0}}{\kappa_{1,1}} U(1)_1 \,
\end{equation}
 remains as an anomaly-free, massless $U(1)$ and can be identified, up to a suitable overall normalisation factor, as $U(1)_Y$.\footnote{{This overall normalisation depends on $\eta_{m_i,n_i}$ and is such that the standard GUT prediction is reproduced in the absence of contributions from the second $E_8$, see the discussion around (\ref{masslessU(1)-gen}).}} This way, the SM hypercharge is now embedded into both $E_8$
 factors. 

The SM matter descends from the massless fermions in the  $\mathbf{248}$ representation of the first $E_8^{(1)}$ upon the breaking (\ref{E81breaking}), with the chiral index determined by the Euler characteristic of certain vector bundles that derive from $V$ and $L_0$. For details we refer to Appendix \ref{App:3gen}, especially Table \ref{tab:SM-spectrum}.   
 Let us instead turn our attention to the fermions charged under the second $E_8$. The $\mathbf{248}$ representation of $E_8^{(2)}$ splits as
\begin{equation}\label{eq:248_splitting} 
    \mathbf{248} \rightarrow \mathbf{133}_0 + \mathbf{1}_0 +\mathbf{56}_1+\mathbf{56}_{-1} +\mathbf{1}_2+\mathbf{1}_{-2} 
\end{equation}
under the breaking (\ref{E82-E7}). 
 The chiral index for these matter states is given by the Euler characteristic
 \begin{equation}
     \chi(L^{q}) = \int_{X} \left(\frac{q^3}{6} c_1^3(L_1) + \frac{q}{12} c_1(L_1) \wedge  c_2(X)\right) \,,
 \end{equation}
where $q$ denotes the $U(1)_1$ charge of the state, $c_1(L_1)$ is the field strength of the line bundle $L_1$, and $c_2(X)$ is the second Chern class of the tangent bundle on $X$. 
 For this type of embedding, and $U(1)_0$ normalised as in Table \ref{tab:SM-spectrum} of Appendix \ref{App:3gen}, the group theory factors are computed as \cite{Blumenhagen:2006ux} 
\begin{equation}\label{kappaeta-example1}
    \kappa_{0,0} = 12, \qquad \eta_{0,0} = 60, \qquad \kappa_{1,1} = -4, \qquad \eta_{1,1} = 4 \,.
\end{equation}
This fixes the massless linear combination (\ref{lin-comb1}) of $U(1)$s to be 
proportional to $U(1)_0 + 3 U(1)_1$.

Only $\mathbf{56}_1$ and $\mathbf{1}_{-2}$ can in principle appear in the charged massless spectrum 
since $\mathbf{133}_0$ and $\mathbf{1}_0$ have no charge under the $U(1)_1$. However, such massless particles would lead to a phenomenological disaster as they carry hypercharge and hence constitute exotic matter. To avoid chiral massless fermions from $\mathbf{56}_1$ and $\mathbf{1}_{-2}$ one must therefore impose the constraints
\begin{equation}
    \chi(L_1^{q_i})=0\,,  \qquad q_i = 1, 2  \,
\end{equation}
on the gauge background. In \cite{Blumenhagen:2006ux} examples of gauge backgrounds  with this property have been found that in addition give rise to three chiral generations of SM matter, and satisfy the D-term condition as well as the Bianchi identity.

The above type of construction 
exemplifies a way to embed 
 hypercharge in a non-standard manner into both $E_8$ factors while at the same time avoiding chiral fermions with exotic charges.
  As we will discuss in more detail in Sections \ref{sec_varphi} and \ref{sec_theta2varphi}, in such models there are two candidates for a light ALP coupling to hypercharge, namely the universal combination $\theta_2$ defined in (\ref{eq:4dEFT_axion_couplings}), coupling to all gauge groups from $E_8^{(2)}$, and an additional linear combination of axions, $\varphi$, that does not couple to the non-abelian gauge group factors from $E_8^{(2)}$.
  Due to the appearance of the confining hidden $E_7$ gauge group factor (see~\eqref{E82-E7}),  $\theta_2$ does not give rise to a light ALP. Indeed, the axion mass will be approximately given by \begin{equation}\label{E7mass}
 m_{\theta_{2}}\sim {\Lambda_{E_7}^2}/{f_{\theta_2}} \,,
 \end{equation}
where the confinement scale of $E_7$ is expected to lie in the region $\Lambda_{E_7}\gtrsim 10^{12}$ GeV.
This leads to a coupling-to-mass ratio ${g_{\theta_{2}\gamma}}\lesssim 10^{-27} \text{GeV}^{-2}$. Since this is many orders of magnitude smaller than the QCD axion prediction, the axion $\theta_2$ easily satisfies the bound~\eqref{eq:coupling_mass_relation}. 
 A more systematic analysis of this effect for other confining gauge group factors will be given in Section \ref{sec:IR_instanton}.
 On the other hand, $\theta_1$ couples to gluons and lies on top of the standard QCD axion band in the $(g_{a\gamma},m_a)$ plane. Finally, the mass for the second ALP, $\varphi$, is independent of strong gauge dynamics and will be determined later. 
 
Note furthermore that the unification of gauge couplings and the standard $\sin\theta_w^2$ prediction from unification is modified at tree-level but receives threshold corrections that can help to reproduce the measured values~\cite{Blumenhagen:2006ux}. We will come back to this important point in Section \ref{sec:unif_gauge_coupling}.
 
\subsection{Non-standard hypercharge embedding without hidden non-abelian sectors}\label{sec:constrain_line_bundle}
Models without a large axion mass of the form (\ref{E7mass}) for $\theta_2$ originating from confining hidden sector gauge dynamics require that the second $E_8$ is broken only to
$U(1)$ factors. Such models represent a qualitatively different type of scenario from the point of axion physics. 

A simple way to avoid confining gauge dynamics in the hidden sector is 
 to break 
 \begin{equation} \label{hiddenpattern}
E_8^{(2)} \to \prod_{m_2=1}^8 U(1)_{m_2}:  \qquad {\bf 248} \to \sum_i \mathbf{1}_{\vec{q_i} }
\end{equation}
through a direct sum of line bundles\footnote{Alternatively, one can consider a gauge background $W_2 = \oplus_{m_2=1}^n L_{m_2} \oplus V_2$ with $c_1(W_2) =0$ including both line bundles and a non-abelian gauge bundle $V_2$ (which in turn may split into a direct sum). This reduces the number of unbroken $U(1)$ factors. Such backgrounds can be obtained, for example, by deforming a subset of the line bundles in (\ref{linebundles-a}) into one or several non-split extension bundles with a non-abelian structure group which then give rise to $V_2$.}
\begin{equation} \label{linebundles-a}
W_2= \oplus_{m_2=1}^8 L_{m_2} \,, \qquad c_1(W_2)=0 \,.
\end{equation}
This background can be combined with a gauge bundle breaking the first $E_8$
 to $SU(3) \times SU(2) \times U(1)_0$ such that a linear combination of $U(1)$s
 remains massless and can be identified with the SM hypercharge. 
 
 All singlet states from $E_8^{(2)}$ with charge under this $U(1)_Y$ must have a vanishing chiral index.
  As we explain in Appendix \ref{App:3gen}, this is most easily ensured if the Calabi-Yau threefold $X$ is elliptically fibered over a complex surface $B_2$ and if the first Chern classes of the line bundles $L_{m_2}$ come entirely 
 from $B_2$.
These conditions must be satisfied together with the D-term condition, the Bianchi identity and while at the same time obtaining three chiral generations of SM matter from the first $E_8$ gauge group. As proof of principle, we present an example fulfilling these constraints in Appendix \ref{App:3gen}.
 Finally, note that even if the chiral index for all $U(1)_{m_2}$ charged matter states vanishes, there can appear massless vectorlike pairs $(\phi_j, \tilde \phi_j)$ of charged ${\cal N}=1$ chiral superfields. 
 After supersymmetry breaking, these vector-like pairs are expected to acquire a mass because they are not protected by any symmetry. In phenomenologically viable models, this must be the case in particular for all fields carrying hypercharge. 

\subsection{Unification of gauge couplings and stringy threshold corrections}\label{sec:unif_gauge_coupling}
 In this section we analyse gauge coupling unification  in minimal models of non-standard hypercharge embedding.
 The gauge couplings of both $E_8$ factors are equal at tree-level and receive one-loop threshold corrections from heavy string modes. These depend on the K\"ahler moduli of $X$ as well as on the moduli of the spacetime-filling NS5-branes appearing the Bianchi identity \eqref{Biachni-Id-gen} \cite{Derendinger:1985cv,Ibanez:1986xy,Choi:1985bz,Nilles:1997vk,Stieberger:1998yi}.
  In the regime of large string coupling, the latter can be interpreted as the position moduli of the 5-branes along the eleventh dimension of heterotic M-theory \cite{Lukas:1997fg,Lukas:1998ew,Carlevaro:2005bk}.
 The threshold corrections are universal for each non-abelian gauge group that originates from the same $E_8$.  On the other hand, $U(1)$ gauge sectors receive additional corrections which are non-universal and depend on the details of the line bundles~\cite{Blumenhagen:2005ga,Blumenhagen:2006ux}. 
 These non-universal corrections also have an impact for axion physics, as they introduce a new linear combination of axions that couples to photons without coupling to any YM group. This will be shown in Section \ref{sec_varphi}.

The threshold corrections to the non-abelian gauge couplings can be parametrised as
\begin{align}\label{eq:YM_corrected_with_threshold}
    &\alpha_{{\rm YM}, 0}^{-1}=\alpha_{\rm tree}^{-1}+\Delta_0^{\rm YM}\,,\\&
        \alpha_{{\rm YM}, 1}^{-1}=\alpha_{\rm tree}^{-1}+\Delta_1^{\rm YM}\,, \label{eq:YM_corrected_with_threshold-2}
\end{align}
with
\begin{equation}
    \Delta_{0,1}^{\rm YM}=-\frac{1}{l_s^2}\int_X J\wedge (\tr_{1,2}\bar F^2-\frac{1}{2}\tr \bar R^2 - (\frac{1}{2} \pm \lambda_5)^2 [W])\,.
\end{equation}
Here $[W]$ denotes the class of the curve on $X$ wrapped by spacetime-filling NS5-branes entering the Bianchi identity (\ref{Biachni-Id-gen}) and $0 \leq \lambda_5 \leq \frac{1}{2}$ parametrises the position of the 5-brane relative to the middle of the interval associated with the eleventh dimension.\footnote{We are assuming here for simplicity that the NS5-branes wrap a single curve.}
From the Bianchi identity it follows that 
\begin{equation}
\Delta_0^{\rm YM}+\Delta_1^{\rm YM} = - (\frac{1}{2} - \lambda_5^2)  \int_X J \wedge [W]\leq 0 \,.
\end{equation}

The gauge kinetic functions of the abelian gauge group factors receive additional non-universal threshold corrections. For a gauge group factor $U(1)_{m_1}$ and $U(1)_{m_2}$ embedded into the first and second $E_8$, respectively, the real part of the gauge kinetic function reads~\cite{Blumenhagen:2005ga,Blumenhagen:2006ux}
\begin{eqnarray} \label{f11def}
  \alpha^{-1}_{m_i n_i} = \delta_{m_i, n_i}\frac{\eta_{m_i,m_i}}{4} \alpha_{{\rm YM},i}^{-1} + \Delta_{m_i,n_i}  \,, 
\end{eqnarray}
where the non-universal threshold corrections are defined as 
\begin{eqnarray} \label{Detltami-gen}
\Delta_{m_i, n_i} &=& -\frac{\kappa_{m_i,m_i} \kappa_{n_i,n_i} }{12}  \int_X J \wedge c_1(L_{m_i})\wedge c_1(L_{n_i}) \,, \\ \quad \Delta_{m_1,n_2} &=& \frac{\kappa_{m_1,m_1}\kappa_{n_2,n_2} }{24}  \int_X J \wedge c_1(L_{m_1}) \wedge c_1(L_{n_2})\,.  
\end{eqnarray}
These expressions depend on the group theoretical invariants $\eta_{m_i,m_i}$ and $\kappa_{m_i,n_j}$ defined in (\ref{traces-inv.}).

Let us note for later purposes that there is always one distinguished linear combination of $U(1)_{m_2}$ factors from $E_8^{(2)}$, given by 
\begin{equation} \label{def-U1D}
U(1)_D = 
 N^{-1}_D \sum_{m_2}\kappa^{-1}_{m_2,m_2} U(1)_{m_2} \,, \qquad N_D = \sqrt{\sum_{m_2} \frac{\eta_{m_2,m_2}}{4} \kappa_{m_2,m_2}^{-2}} \,,
\end{equation}
 for which the non-universal part of the threshold corrections vanishes. 
This follows from the explicit expression (\ref{Detltami-gen}) for the thesholds $\Delta_{m_2, n_2}$ together with the fact that $\sum_{m_2} c_1(L_{m_2}) = 0$ for the type of embeddings under consideration.
Taking into account also the overall normalisation, the square of the inverse gauge coupling of this $U(1)_D$ is then given precisely by 
\begin{equation}
\alpha_D^{-1} = \alpha^{-1}_{\rm YM,1} \,
\end{equation}
as defined in (\ref{eq:YM_corrected_with_threshold-2}).

 Let us define the GUT gauge coupling as the coupling associated to $SU(3)_C$ and $SU(2)_w$ at the GUT scale,
\begin{equation} \label{alphac-def}
\alpha_w=\alpha_C= \alpha_{\rm YM,0} \equiv \alpha_{\rm GUT} \,.
\end{equation}
We can then express the one-loop corrected coupling of the second $E_8$ as
\begin{equation}\label{alphaYM1}
    \alpha_{\rm YM,1}^{-1}=\alpha_{\rm GUT}^{-1}-\Delta_0^{\rm YM}+\Delta_1^{\rm YM}\,.
\end{equation}
Depending on the sign of $-\Delta_0^{\rm YM}+\Delta_1^{\rm YM}$ the gauge coupling in the second $E_8$ is larger or smaller than $\alpha_{\rm GUT}$. This will be crucial later to estimate axion masses from (UV) small gauge instantons.

For simplicity of notation let us assume that $U(1)_Y$ is a linear combination of a $U(1)_0$
from $E_8^{(1)}$ and $U(1)_1$ from $E_8^{(2)}$. This is easily generalised  in the presence of several $U(1)$ factors e.g. from the second $E_8$.
 The gauge coupling at the KK scale receives contributions from both abelian factors and, with the usual SM normalisation,
takes the form
\begin{equation}\label{eq:hypercharge_coupling_with_threshold}
\frac{1}{\alpha_Y}  = \frac{5/3}{\alpha_{\rm GUT}}+ \frac{k_1^{(2)}}{\alpha_{\rm GUT}} + \Delta_Y \,.
\end{equation}
Here $k_1^{(2)}$ is the level of embedding of $U(1)_1$ into the second $E_8$. It depends on the group theoretic embedding and the choice of gauge backgrounds. The threshold correction $\Delta_Y$ depends on $\Delta_1^{\rm YM} - \Delta_0^{\rm YM}$ but also on the non-universal part of the abelian thresholds. 
 Concretely, let us 
 parametrise the two  line bundles associated with $U(1)_0$ and $U(1)_1$
as 
\begin{equation} \label{c1Lirelation}
{c_1(L_i)} = \ell_i  v    \,, \qquad v \in H^2(X, \mathbb Z)   \,, \qquad \ell \in \mathbb Z \,.
\end{equation}
Indeed, both line bundle curvatures must be proportional so that  $U(1)_Y$ is massless, as can be seen from the mass matrix (\ref{massmatrix}), (\ref{massmatrix-1}).
Including the normalisation, the hypercharge is then given by the combination 
\begin{equation} \label{masslessU(1)-gen}
U(1)_Y = \sqrt{\frac{5}{3}} \sqrt{\frac{4}{\eta_{0,0}}} \left (U(1)_0 - \frac{\kappa_{0,0} \ell_0}{\kappa_{1,1} \ell_1} U(1)_1\right )\,.
\end{equation}
We see that in the absence of contributions from $E_8^{(2)}$ we reproduce the standard GUT relation for the gauge couplings.
With the overall normalisation and the factors $\frac{\eta_{i,i}}{4}$ in the gauge couplings (\ref{f11def}) taken into account,
 the level $k_1^{(2)}$ follows as 
\begin{equation} \label{k12-1}
k_1^{(2)} = \frac{5}{3} \left(\frac{\kappa_{0,0} \ell_0}{\kappa_{1,1} \ell_1}\right)^2 \frac{\eta_{1,1}}{\eta_{0,0}} \,,
\end{equation}
and the hypercharge threshold correction takes the form 
\begin{equation}\label{eq:threshold_hypercharge}
\Delta_Y = \Delta_{Y, \rm ab}  + k_1^{(2)} (\Delta_1^{\rm YM} - \Delta_0^{\rm YM}) \,.
\end{equation}
As we will see later, the non-universal contribution $\Delta_{Y, \rm ab}$ is particularly important to determine ALP masses and is given by
\begin{equation}
\Delta_{Y, \rm ab} = \frac{5}{3} \frac{4}{\eta_{0,0}} \left(\Delta_{0,0} + \frac{3}{5}\frac{\eta_{0,0}}{\eta_{1,1}} k_1^{(2)} \Delta_{1,1} - 2 \frac{\kappa_{0,0} \ell_0}{\kappa_{1,1} \ell_1} \Delta_{0,1}\right) = 5 \frac{4}{\eta_{0,0}} \Delta_{0,0} =  5 \widehat{\Delta}_0.
\end{equation}
where in the last step we defined
\begin{equation} \label{def-hatdeltai}
\widehat{\Delta}_i := \frac{4}{\eta_{i,i}} \Delta_{i,i} \,,
\end{equation}

\subsubsection{Unification without charged states at intermediate scales}\label{sec:nocharged_matter_intermedite_scales}

In the absence of new charged matter fields at intermediate energy scales, the standard values for the gauge couplings and the weak mixing angle at low energies are obtained if
\begin{equation} \label{DeltaYalphaGUT}
\Delta_Y = - \frac{k_1^{(2)}}{\alpha_{\rm GUT}} \,.
\end{equation}
 For example, for the model of \cite{Blumenhagen:2006ux} reviewed in Section \ref{sec:Timo's_model}, with group theory factors (\ref{kappaeta-example1}) and $\ell_0 = \ell_1$,
$k_1^{(2)} =1$ and hence the threshold corrections must be sizeable, $\Delta_Y = -{\alpha^{-1}_{\rm GUT}}$.
We noted already that in principle more $U(1)$ factors from the second $E_8$ may contribute to the SM hypercharge in Eq.~\eqref{masslessU(1)-gen} (see also Eq.~\ref{eq:hypercharge_coupling_with_threshold}); in this case, the additional contributions require even larger threshold corrections $\Delta_Y$ or more charged matter. This is disfavored from a theory point of view. In the rest of the paper we consider, for simplicity, scenarios where only one such $U(1)$ factor from the second $E_8$ contributes. Our results can easily be generalised without affecting the conclusions.

By contrast, for $k_1^{(2)} \ll 1$, the standard relation ${\alpha^{-1}_Y} ={\frac{5}{3}}{\alpha^{-1}_{\rm GUT}}$ at the KK scale can be obtained with small threshold corrections, $|\Delta_Y| \ll 1$, and no additional charged states.
Moderately small values of $k_1^{(2)}$ can be engineered, for example, through a suitable parameter $\ell_1$ defined in (\ref{c1Lirelation}).\footnote{Note that in $\Delta^{\rm YM}_1$, the $\ell_1$ dependence on ${\rm tr}_1 F^2$  cancels the $\ell_1$ dependence of $k^{(2)}_1$.} 
An example with $\ell_0 = 1$, $\ell_1 = 3$, leading to $k_1^{(2)} = \frac{1}{9}$, is presented in Appendix \ref{app_ksmaller1}.

\subsubsection{Unification with charged states at intermediate scales}\label{sec:charged_matter_intermedite_scales}
Let us now discuss how new charged states at intermediate scales can improve the unification of gauge couplings.
We assume momentarily that we are in a region in moduli space where the threshold corrections $\Delta_Y$ appearing in (\ref{eq:hypercharge_coupling_with_threshold}) are negligible and we are left with the matching condition $\alpha_Y^{-1}=(5/3+k_1^{(2)})\alpha_{\rm GUT}^{-1}$. Assuming that $k_1^{(2)}$ is not small, the only way to recover the measured gauge couplings at low energies is by having new vector-like charged states $\Psi_i$ at intermediate scales $m_{\rm EW}\lesssim M_{\Psi_i} \lesssim M_{\rm GUT}$. 

Adding extra matter charged only under $U(1)_Y$ will not improve the situation since it will tend to make the hypercharge coupling weaker in the IR, which is the opposite of what Eq.~\eqref{eq:hypercharge_coupling_with_threshold} requires. To recover the observed values of the gauge couplings and $\sin^2\theta_w$ there should exist extra matter beyond the MSSM spectrum with $SU(3)_C$ and $SU(2)_w$ charges.
These new particles change the $\beta$-functions in a way that leads to the observed IR gauge couplings only if we have a stronger unified gauge coupling than the standard MSSM prediction, defined as $\alpha_2=\alpha_3=\alpha_{\rm GUT}$.
Due to the new states, the unified gauge coupling shifts with respect to the initial MSSM prediction (that is, without the new charged states), defined as $\alpha_{\rm initial}$,
\begin{equation}\label{eq:gauge_coupling_light_matter}
    \alpha_{\rm GUT}^{-1}=\alpha_{\rm initial}^{-1}+\Delta \alpha^{-1}\,.
\end{equation}
Recall that for low-energy supersymmetry, $\alpha_{\rm initial}^{-1}=24$.  The induced shift for $n_F$ superfields is negative and can be expressed as
\begin{align}\label{eq:shift_GUT_gauge_coupling}
	\Delta \alpha^{-1}
	= - \sum_{R}\frac{n_F T_R}{\pi}\ln\left(  M_{\rm GUT} / M_\Psi \right) \,,\,\,\, \text{for }i=2,3\,,
\end{align}
with $T_{R}$ being the Dynkin index of the $R$ representation.  

We note, however, that the presence of new charged states alone does generically not solve the issue of gauge coupling unification. This would require intermediate charged states with only weak and/or color charges. 
These states should be part of the predicted non-chiral spectrum of massless particles. However, in the simply connected models that we study, generically no representation with $SU(2)_w$, $SU(3)_C$ charge appears without also carrying hypercharge. This can be seen, for example, from Table \ref{tab:SM-spectrum}. 
In this case, the beta function of hypercharge is necessarily modified above the mass scale of the new degrees of freedom as well. One could add, for example, multiple copies of new (vector-like) superfields transforming as $(1,2,-1/2)$ or $(3,1,-1/3)$, which are the ones with smallest impact in the hypercharge beta function, $\beta_Y$. However, a careful analysis reveals that in this case one can only obtain the observed gauge couplings if $SU(2)$ and $SU(3)$ unify (defining the GUT scale, $M_{\rm GUT}$) well below the usual lower bound obtained from proton decay searches~\cite{Super-Kamiokande:2020wjk}. Similar to the discussion in in~\cite{Blumenhagen:2006ux}, a simple example is the case where there exist two superfields transforming as $(1,2,-1/2)$ around the TeV scale. In this case, $M_{\rm GUT}=4\times 10^{11}$ GeV and $\alpha_{\rm GUT}^{-1}=19$, in conflict with the required proton stability.

We cannot exclude the possibility that in models with more complicated embeddings, with multiple $U(1)$s contributing to hypercharge, there exist new charged states that are in the end uncharged under $U(1)_Y$, although this seems to be contrived. Similarly, it is conceivable that new charged states (with $SU(2),SU(3)$ quantum numbers as well as hypercharge), {\it together} with the threshold corrections studied above, result in the observed gauge coupling values. In either case, the fact that $\alpha_{\rm GUT}^{-1}$ is smaller than the minimal predictions of the MSSM (or related UV completions of the SM) has crucial implications for the ALP and the QCD axion that will be studied later in sections \ref{sec:small_gs} and \ref{sec:intermediate_states_ALP_potential}.

\subsection{Line bundles, threshold corrections, and axion couplings}
\label{sec_varphi}
The breaking of the symmetry $E_8^{(i)}\rightarrow G_i\times \prod_{m_i} U(1)_{m_i} $ by topologically non-trivial line bundles has important implications for axion couplings, too. In Eq.~\eqref{Detltami-gen} we have seen that it generates non-universal threshold corrections -- that is, corrections to the gauge coupling which are different for each of the $U(1)_{m_i}$. Since these corrections are holomorphic, they also shift the model dependent axion couplings to the unbroken $U(1)$ gauge groups. In this section we show that when $U(1)_Y$ arises as a linear combination of $U(1)$s via embeddings as studied so far, a new linear combination of axions different from the combinations $\theta_{1,2}$ in (\ref{eq:linear_comb_theta_12}) appears and couples to photons via the anomaly.

The corrections we consider originate  from cross-terms 
in the GS counter-terms \eqref{eq:GS} between gauge field strengths tangent to Minkowski and the internal space \cite{Blumenhagen:2005ga},
\begin{equation}
 S_{\rm GS}\supset \frac{1}{768\pi^3} \int B_2\wedge \tr_i(F\bar F)^2- \frac{1}{768\pi^3}\int B_2\wedge \tr_1(F\bar F)\tr_2(F\bar F)\,.
\end{equation}
These pieces of the GS counter-terms are only phenomenologically relevant for the $U(1)$ symmetries that contribute to hypercharge, which as in Section \ref{sec:unif_gauge_coupling}  we parametrise as a linear combination of $U(1)_0$ and $U(1)_1$ from the two different $E_8$ factors. 
 Since the associated line bundles are proportional, $c_1(L_i)=\ell_i v$, $i=0,1$, to ensure that  hypercharge remains massless, it can be shown that the same axion linear combination couples to the $U(1)_i$ that participate in hypercharge. This linear combination only involves model dependent axions and is given by 
\begin{equation}
    \varphi=  \sum_j b_j\int\beta_j\wedge v^2 = \sum_j \tilde n_j b_j\,
\end{equation}
with $\beta_j$ a basis of $H^2(X)$
into which the 2-form $B_2$ is expanded to yield the model dependent axions, $B_2 = \sum_j b_j \beta_j + \ldots$.
This axion obtains a topological coupling to the $U(1)_0$ and $U(1)_1$ gauge bosons,
\begin{equation}\label{eq:varphi_coupling_to_gaugebosons}
    \varphi \left ( \kappa_{0,0}^2 \ell_0^2 F_0 \tilde F_0 + \kappa_{1,1}^2 \ell_1^2 F_1 \tilde F_1 - \kappa_{0,0}\kappa_{1,1} \ell_0 \ell_1 F_0 \tilde F_1\right )\,.
\end{equation}
Phenomenologically, this is important 
  because $\varphi$ is a new linear combination  different from $\theta_1$ and $\theta_2$ defined in (\ref{eq:linear_comb_theta_12}), which couple universally to all the gauge bosons that survive in 4d from the first and second $E_8$, respectively. 

The coupling of $\varphi$ to hypercharge can be obtained from~\eqref{eq:varphi_coupling_to_gaugebosons}. Equivalently, one can invoke holomorphy and obtain it from the non-universal hypercharge shift, calculated in Eq.~\eqref{eq:threshold_hypercharge}. This leads to the coupling
\begin{equation}\label{eq:varphi_coupling_to_gaugebosons2}
  \Delta\mathcal{L}=  \frac{\varphi}{8\pi^2}\frac{(\kappa_{0,0}\ell_0)^2}{\eta_{0,0}} k_1^{(1)}F_Y\tilde F_Y \,.
\end{equation}
Similar to the hypercharge threshold corrections, this coupling  vanishes as $(\kappa_{0,0}\ell_0)^2\rightarrow 0$ or as $g_s^2 \to 0$. The latter can be seen from the fact that the decay constant for model dependent axions grows as $g_s$ decreases~\cite{Svrcek:2006yi}. Unlike the axions $\theta_{1,2}$, $\varphi$ does not couple to YM sectors, meaning that its mass can only be generated by UV physics. Indeed, the mass of $\varphi$ will be induced by worldsheet instantons as well as small gauge instantons and will be derived in later sections.

All in all, we see that when hypercharge is obtained as a result of gauge symmetry breaking via non-trivial line bundles, a new axion linear combination can appear and couple to photons without obtaining a potential from QCD. Remarkably, $\varphi$ remains light even if there is a confining interaction $G_{\rm hidden}$ from $E^{(2)}_8$. In this case, $\theta_2$ will gain a large mass (see the discussion around \eqref{E7mass}), but $\varphi$ will be unaffected. In later sections we will study the different shift-symmetry breaking effects for $\varphi$. 

\subsection{The QCD axion and the ALP(s) in models with non-standard embedding} \label{sec_theta2varphi}
Let us now study the axion couplings in a heterotic model with non-standard embedding as constructed above, where the groups $SU(3)_C$ and $SU(2)_w$ are embedded in the first $E_8$ with no contribution from the second factor, while $U(1)_Y$ is embedded in both $E_8$ factors.  Following~\cite{Agrawal:2024ejr}, the axion couplings in the 4d EFT are obtained from the GS counter-term after dimensionally reducing the 10d SUGRA action. Crucially, these couplings are quantized, meaning that they are not affected by renormalization group flow. The only model dependence of the axion-gauge boson coupling comes from the embedding of the SM gauge group into the 10d group and the line bundle, which are classified by a small set of integers. 

Below the EWSB scale there are different linear combinations of axions that couple to gauge bosons via the anomaly. These are $\theta_1$ and $\theta_2$, which couple \textit{universally} to the gauge bosons arising from each $E_8$ factor, and $\varphi$, which, as shown in Section (\ref{sec_varphi}),  arises after taking into account non-universal threshold corrections. In the class of theories we consider, these axions couple to gluons and photons as  
\begin{align}\label{eq:eff_Lag_non-standard_embedding}
    \mathcal{L}&= \frac{\theta_1}{{8\pi^2}} \left[ k^{(1)}_3 \alpha_3 \tr G\tilde{G} +   \left( \frac{(k^{(1)}_1)^2}{k^{(1)}_1 + k^{(2)}_1} + k^{(1)}_2 \right)\alpha_\EM F_\EM  \tilde{F}_\EM \right] 
    \\& + \frac{\theta_2}{{8\pi^2}} \left( \frac{(k^{(2)}_1)^2}{k^{(1)}_1 + k^{(2)}_1}  \right)\alpha_\EM F_\EM  \tilde{F}_\EM 
    +  \frac{\varphi}{{8\pi^2}}\frac{(\kappa_{0,0}\ell_0)^2}{\eta_{0,0}} k_1^{(1)} \alpha_\EM F_\EM\tilde F_\EM \,.
\end{align}
 Note that $\theta_i$ and $\varphi$ are dimensionless. To compute the dimensionful coupling to photons, $g_{a\gamma}$, of the canonically normalized, dimensionful axions one has to determine the effective decay constant for each linear combination, see~\cite{Agrawal:2024ejr,Leedom:2025mlr}. 
Here $ k^{(1)}_3, \,  k^{(1)}_2, \,  k^{(1)}_1$ are the embedding levels of the SM gauge groups in  $E_8^{(1)}$ and $k^{(2)}_1$ is the embedding level of hypercharge into the second $E_8^{(2)}$. 
Since all SM fermions come from gauginos transforming as $(\mathbf{248},\mathbf{1})$, the normalisation of the relevant $U(1)$ generators is $k_1^{(1)}=5/3$, 
and $k_1^{(2)}$ is in general given by Eq.~(\ref{k12-1}). {This expression relates $k_1^{(2)}$ to the group theoretic factors $\eta_{ii}$ and the integers parameterizing the line bundle in Eq.~\eqref{c1Lirelation}.}

On the other hand we will assume that $SU(3)_C$ and $SU(2)_w$ are embedded at level one into $E_8^{(1)}$, $k_3^{(1)}=k_2^{(1)}=1$. 
We note that this is actually a conservative choice. A higher level of embedding introduces (fractional) UV gauge instantons with a smaller action, $S_{\rm UV}\sim \frac{2\pi}{k\alpha_{\rm GUT}}$. For example, if the EW or QCD groups arise as a diagonal subgroup of $SU(N)\times SU(N)$, where each $SU(N)$ factor is embedded at level one into a different $E_8$, one has an effective level-two embedding, $k_{2,3}=k_{2,3}^{(1)}+k_{2,3}^{(2)}=2$. As will be discussed later in Sections \ref{sec:ALP_potential} and \ref{sec_thresholds}, this would have two main effects. First, it would make the ALPs $\theta_2$, $\varphi$ heavier  compared to the estimates that we will obtain. Second, a non-trivial embedding of QCD can spoil the QCD axion solution to the strong CP problem even for $k_3=2$; a similar situation will be discussed in Section \ref{sec:intermediate_states_ALP_potential}.

In the absence of sizeable effects from other instantons, the linear combination $\theta_1$ obtains its mass from the coupling to gluons and behaves as the QCD axion. {Its coupling-to-mass ratio $g_{\theta_1\gamma}/m_{\theta_1}$ is the standard one for a QCD axion except that the ratio of anomaly coefficients $E/N$, with $E$ ($N$) being the anomaly coefficient with QED (QCD), is different with respect to minimal 4d GUTs, which predict $E/N=(k_1+k_2)/k_3$~\cite{Agrawal:2022lsp}. In the class of models we consider, once the non-trivial embedding into the second $E_8$ is taken into account, the ratio for $\theta_1$ is given by
\begin{equation}
    \frac{E}{N}= \frac{\frac{(k_1^{(1)})^2}{k_1^{(1)}+k_1^{(2)}}+k_2^{(1)}}{k_3^{(1)}} \,.
\end{equation}
}

{On the other hand, $\theta_2$ and $\varphi$ do not couple to gluons. 
The coupling of $\theta_2$ to photons, $g_{\theta_2\gamma}$, is proportional to the effective 4d anomaly coefficient with QED,
\begin{equation}
    \mathcal A_{\theta_2}= \frac{(k^{(2)}_1)^2}{k^{(1)}_1 + k^{(2)}_1}\,,
\end{equation}
and its mass $m_{\theta_2}$ depends on the unbroken gauge group from the second $E_8$. As expected, as $k_1^{(2)}\rightarrow 0$, $\theta_2$ is decoupled from photons.
In the next section we will compute $g_{\theta_2\gamma}/m_{\theta_2}$ for different unbroken gauge groups in the case $k_1^{(2)}\neq 0$.} 

{The case of $\varphi$ is slightly different. Its coupling to photons is induced due to the origin of $U(1)_Y$ as a massless linear combination of the $U(1)$s left unbroken by the line bundles (see Eq.~\eqref{masslessU(1)-gen}). This can be seen from its effective anomaly coefficient $\mathcal{A}_\varphi=\frac{(\kappa_{0,0}\ell_0)^2}{\eta_{0,0}} k_1^{(1)}$, which tells us that only as we turn off the line bundle in the first $E_8$ does the coupling of $\varphi$ to photons vanish, too. This anomaly coefficient $\mathcal{A}_\varphi$ decreases at the same rate as the non-universal threshold correction to hypercharge $\Delta_{Y,\rm{ab}}$, again due to holomorphy.}

We can now estimate the coupling to photons for the QCD axion, $\theta_1$, which depends on the value of $k_1^{(2)}$.
For example, 
for $k_1^{(1)}=5/3$, $k_1^{(2)}=1$, and $k^{(1)}_2=k^{(1)}_3=1$, as in the concrete examples 
 of Sections \ref{sec:Timo's_model} and \ref{sec:constrain_line_bundle},
the ratio of anomaly coefficients is
\begin{equation}
    \frac{E}{N}=\frac{49}{24}\,.
\end{equation}
This prediction is particularly important phenomenologically. The reason is that due to a partial cancellation with the contribution from the pion mixing (see Eq.~\eqref{eq:coupling_mass_relation}), this value of $E/N$ leads to a ratio
\begin{equation}
    g_{\theta_1\gamma}/m_{\theta_1}\approx 0.01\text{ GeV}^{-2}\,,
\end{equation}
which is a factor $\sim 6$ smaller than the QCD axion coupling to photons in the standard GUT case. This cancellation is crucial for experimental searches, since it demonstrates that the QCD line in these models with non-standard hypercharge embedding is harder to access experimentally (see  Figure~\ref{fig:ParameterSpace}).

\section{The irreducible ALP potential}\label{sec:ALP_potential}
From a 4d EFT point of view, the potential for the axions $\theta_2$ {and $\varphi$} is of the form
\begin{equation} \label{eq:theta2-pot}
    V (\theta_2,\varphi) = -\kappa_{\theta_2} \Lambda_{\theta_2}^4  \cos \theta_2 - {\kappa_\varphi \Lambda_\varphi^4  \cos \varphi }+ V_{\rm ws}(b_i, \tilde b_r)\,,
\end{equation}
with $V_{\rm ws}(b_i, \tilde b_r)$ being the axion potential from worldsheet instantons. 
The scale $\Lambda_{\theta_2}$ corresponds approximately to the confinement scale of a non-abelian group if the instanton potential is IR dominated. If, on the other hand, no YM sector survives, we will argue that the first part of the potential is generated by 5-brane instantons which can be interpreted as small (UV) gauge instantons. In this case, $\Lambda^4 \approx M_{\rm UV}^4e^{-S}$, with $M_{\rm UV}$ the UV cutoff of the 4d EFT.  
 On the other hand, $\Lambda_\varphi$ can only be generated via small gauge instantons of the $U(1)$ symmetries that participate in the hypercharge linear combination or by worldsheet instantons.
The coefficients $\kappa_{i}$ will be seen to be related to the one-loop determinant; they contain possible chiral suppression due to charged light fermions and low-scale supersymmetry because zero-modes have to be saturated by mass insertions or by interactions with scalar fields.

In this section we study the induced axion potential $V(\theta_2,\varphi)$ for the different possible gauge sectors from the second $E_8$. The first possibility is that there is an unbroken, confining interaction, $G_{\rm hidden}$, that commutes with the $U(1)$ group contributing to hypercharge. In Section \ref{sec:IR_instanton} we argue that in this case, the coupling-to-mass ratio for the  ALP, $g_{\theta_2\gamma}/m_{\theta_2}$, is much smaller than the QCD axion prediction, satisfying the bound \eqref{eq:coupling_mass_relation} in a similar way to the model in Section~\ref{sec:Timo's_model}. 
 The coupling-to-mass ratio for $\varphi$, $g_{\varphi\gamma}/m_{\varphi}$, however, is unaffected.

A  qualitatively different situation occurs when the second $E_8$ is broken to $\prod _{m_2} U(1)_{m_2}$ as in the models of Section \ref{sec:constrain_line_bundle}. 
In Section \ref{sec:SmallInstantons} we  argue that 
in such models, UV instantons generate an ALP potential that can be reliably determined once the value of the unified gauge coupling and the SUSY breaking scale are fixed. We will also consider cases with light fermions in Section \ref{sec:light_fermions},  showing that even though they introduce chiral suppression in $V(\theta_2,\varphi)$, they change the UV value of the gauge coupling and small instantons become more relevant. For this reason, 3-generation MSSM-like constructions offer a reliable upper bound to $g_{a\gamma}/m_a$ 
 for both ALPs, $\theta_2$ and $\varphi$, which are comparable in the limit of small threshold corrections. The leading order estimates for different values of $\alpha_{\rm GUT}$ are shown in  Figure \ref{fig:ParameterSpace}

In Section \ref{sec:worldsheet_and_ALPs} we estimate the worldsheet instanton potential $V_{\rm ws}(b_i, \tilde b_r)$. We will see that in certain situations they dominate in the axion potential over small gauge instantons. In this case, for some of the model dependent axions $b_i$, we expect $|V_{\rm ws}(b_i)|\gg \kappa \Lambda^4_{\theta_2,\varphi}$, which in turn implies that such $b_i$ can be integrated out and do not affect the low-energy phenomenology. In the most extreme scenarios this results {into the linear combination $\varphi$ being integrated out completely, and $\theta_2$ being given only by the model independent axion, $\theta_2 = a$}. This is very important phenomenologically, since it  automatically leads to a situation where Eq.~\eqref{eq:coupling_mass_relation} is satisfied regardless of the non-standard hypercharge embedding.

Throughout this section we will, for ease of presentation, estimate the ALP potential $V(\theta_2,\varphi)$ at leading order -- that is, under the assumption of negligible stringy threshold corrections ($|\Delta_Y| \ll 1$ in (\ref{eq:hypercharge_coupling_with_threshold})) and, unless explicitly stated (as in Section~\ref{sec:light_fermions}), assuming only 3-generations of SM (MSSM) matter fields and no additional light fields charged under the SM. As noted in Section~\ref{sec:unif_gauge_coupling}, these models only have a chance to be realistic if $k_1^{(2)} \ll 1$. However, they offer the advantage that an upper bound $g_{a\gamma}/m_a$ can easily be obtained as a function of the supersymmetry breaking scale.  After gaining intuition about the shift symmetry breaking effects and how they induce an ALP potential we will further refine the estimates to the coupling-to-mass ratio for the ALPs in more realistic scenarios, including the effect of threshold corrections, in Section~\ref{sec_thresholds}. Recall that, strictly speaking, the coupling of $\varphi$ to gauge bosons vanishes as $\Delta_Y\rightarrow 0$. However, in preparation for Section \ref{sec_thresholds}, in this section we will consider the non-perturbative effects that generate a mass for this axion. 

\subsection{ALP potential from IR YM instantons}\label{sec:IR_instanton}
In this section we analyse models where the second $E_8$ is broken  to
\begin{equation}
    U(1)^k\times G_{\rm hidden} \,,
\end{equation}
{with $G_{\rm hidden}$ a YM sector that generates a mass for $\theta_2$, but not for $\varphi$, as exemplified in Section \ref{sec:Timo's_model}}. 
 There are two qualitatively different classes of such models to consider.

\subsubsection{Confining $G_{\rm hidden}$ with massive fermions}\label{sec:confining_group_massive_fermions}
In the first type of models, there are no massless fermions charged under $G_{\rm hidden}$.
Let us assume for simplicity that $G_{\rm hidden}$ is a non-abelian hidden sector that admits real representations. Matter transforming in such representations is vector-like and hence expected to gain a mass, at the very least after supersymmetry breaking. Depending on the mass of the fermions, $G_{\rm hidden}$ then resembles either a pure SYM sector or QCD with a number of light quarks. 

Once the spectrum is known, one can compute the beta function for $G_{\rm hidden}$ and estimate  the scale at which the gauge coupling becomes strong, starting from the boundary condition $\alpha_h(R^{-1})\sim \alpha_{\rm GUT}=1/24$ at the KK scale. Solving the RG equation for the gauge coupling allows a rough estimate of the strong coupling scale, $\Lambda_{\rm hidden}$, that enters the potential (\ref{eq:theta2-pot}).
For example, for a supersymmetric YM theory, one has 
\begin{equation}
    \alpha^{-1} (\mu) = \alpha^{-1}(M) - \frac{3C_2}{2\pi} \ln \frac{M}{\mu}\,.
\end{equation}
For the smallest non-abelian group, $SU(2)$, with only gauginos, we have $C_2=2$, leading to
\begin{equation}
    \Lambda_{SU(2)}\approx 10^5 \text{ GeV}\,.
\end{equation}
Other possible groups, assuming a pure supersymmetric YM sector, lead to the following strong coupling scales
\begin{align}
    &SU(3): \Lambda_{SU(3)}\approx 10^8 \text{GeV}\,, \,\,\,\,\,\,
    SU(4): \Lambda_{SU(4)}\approx 10^{10} \text{GeV} \,.
\end{align}
For the level of precision that we require, these approximate estimates are sufficient.

In the presence of low-energy SUSY, $m_{\rm susy}\ll \Lambda_{SU(N)}$, gauginos induce a mild chiral suppression and the axion potential becomes
\begin{equation}
    V(\theta_2)\sim - m_{\rm susy} \Lambda_{SU(N)}^3 \cos \theta_2 \,.
\end{equation}
From this, one can estimate the ratio of the axion-photon coupling to its mass, 
\begin{equation}
\frac{g_{\theta_2\gamma}}{m_{\theta_2}}\sim \frac{\alpha_{\rm em}}{2\pi}\frac{1}{\sqrt{m_{susy}\Lambda^3_{SU(N)}}}   \,.
\end{equation}
For the values above, we see that the axion $\theta_2$ easily satisfies the relation \eqref{eq:coupling_mass_relation}.
The result for different groups is shown in  Figure \ref{fig:ParameterSpace}.
We conclude that in situations where the gauge background leaves an unbroken YM sector, $\theta_2$ has a coupling-to-mass ratio much smaller than the QCD axion prediction. If produced in the early universe (e.g. via misalignment mechanism), $\theta_2$ could have an interesting impact in cosmology due to its decay into photons and other SM particles.

At this point it appears that only  $\theta_2$ gained a large mass but $\varphi$ still behaves as a light ALP with $g_{\varphi\gamma}/m_\varphi$  larger than the QCD axion prediction. However, in the next section we will see that independently of strong gauge dynamics,
the linear combination $\varphi$ also gains a large mass through UV instantons.

\subsubsection{Confining $G_{\rm hidden}$ with massless fermions}\label{sec:massless_fermions}
We now come to the second class of models,  with some massless fermion,
$\psi$, charged under $G_{\rm hidden}$.
 This can either be a limiting case of the situation analysed in the previous section, with $m_\psi = 0$, or correspond to a massless chiral fermion in models with non-vanishing chiral index.\footnote{The latter corresponds to a confining chiral gauge theory. Such models are less well understood than theories with vector-like confining interactions, see however \cite{Csaki:2021xhi,Goh:2025oes}.}

In the presence of a massless charged fermion, there is an additional $U(1)_\psi$ global symmetry associated to chiral rotations of $\psi$. This symmetry has a mixed $[G_{\rm hidden}]^2\times U(1)_\psi$ anomaly. A chiral rotation of the fermion removes the coupling of the axion to the confining interaction, $\theta_2 G_h\tilde G_h$, at the cost of introducing derivative couplings to the fermion $\psi$,
\begin{equation}
    {\partial_\mu \theta_2}\psi\gamma^\mu \gamma^5\psi\,.
\end{equation}
As a result, there is no light axion $\theta_2$ coupled to gauge bosons from the second $E_8$ (including photons) and, instead, just a massless Goldstone boson derivatively coupled to the hadronic states of the confining interaction $G_{\rm hidden}$. 

If due to the strongly coupled sector $G_{\rm hidden}$ the fermion $\psi$ forms a condensate $\langle \psi \psi \rangle$, the anomalous symmetry $U(1)_\psi$ is broken spontaneously at around the strong coupling scale. {In this case the ALP $\theta_2$ becomes an $\eta^\prime$-like state that is coupled to $G_h\tilde G_h$ of the hidden confining group and gains a large mass close to the scale of the condensate.} This resembles scenarios where one of the SM quarks is massless and the strong CP problem is solved by the $\eta^\prime$ meson.

\subsection{ALP potential from small (UV) gauge instantons}\label{sec:SmallInstantons}
Let us now estimate the ALP potential that is generated independently of strong gauge dynamics. This potential is relevant both for $\theta_2$ and $\varphi$. We will consider supersymmetric compactifications with $m_{\rm susy}\ll M_{\rm GUT}$ although the results can be generalised to situations without low-energy supersymmetry. 

Suppose the gauge background breaks $E_8^{(2)}$ to $\prod_{m} U(1)_m (\times G_{\rm hidden})$.\footnote{To avoid clutter of notation we henceforth drop the subindex in the label $m_2$ for the $U(1)$ factors from the second $E_8$ whenever the context is clear.}   Irrespective of the presence of a strongly coupled YM sector, Euclidean heterotic 5-branes wrapping the entire compact space $X$, fully localised in Minkowski, generate a potential which is equivalent to small gauge instantons at the compactification scale \cite{Witten:1995gx}. 
 For brevity, we will ignore the hidden YM sector from now on, keeping in mind that the discussion immediately generalises accordingly in its presence.
 
As we will argue below, (fluxed) 5-brane instantons contribute to the superpotential terms of the schematic form 
\begin{equation}\label{eq:superpotential_contribution_axion_potential}
    W \supset \sum_{m \in \cal I}  A_m \left ( \prod_l (\phi_l \tilde \phi_l)^{n_{m,l}} \right ) \, e^{- \frac{8\pi^2}{g_m^2} + i f_m(a,b_i,\tilde b_r)}\,.
\end{equation}
Here we are focusing, for simplicity, on the terms associated with those  $U(1)_m$ gauge factors for $m\in {\cal I}$ which contribute to the linear combination associated with hypercharge $U(1)_Y$. The reason is that these cannot admit chiral charged matter,
but at best vectorlike chiral fields $(\phi_j,\tilde \phi_j)$ charged under these abelian gauge groups (see the discussion at the end of Section \ref{sec:constrain_line_bundle}). Each such pair contributes with a model-dependent multiplicity $n_{m,j}$ to the $m$-th instanton. Furthermore, $A_m$ is a field-independent factor of proportionality and $f_m(a,b_i,\tilde b_r)$ is the imaginary part of the gauge kinetic function of $U(1)_m$. 

Since 5-brane instanton effects in heterotic compactifications are comparatively poorly understood, we refrain from attempting to give a rigorous proof of the appearance of such a superpotential in full generality. However,
its appearance can be argued convincingly in the subclass of heterotic models which are dual to F-theory.\footnote{We refer to \cite{Weigand:2018rez} and references therein for an introduction to F-theory and to \cite{Palti:2020qlc} for a description of the F-theory dual of heterotic 5-brane instantons.} 
 Such heterotic models are formulated on an elliptically fibered Calabi-Yau threefold. 
 The class of models presented in Appendix \ref{App:3gen} in which the chiral index for the hidden exotic matter fields charged under $U(1)_Y$ vanishes is precisely of this type.
 The dual F-theory model is defined on an elliptic Calabi-Yau 4-fold over a base $B_3$, which  in turn is a $\mathbb P^1$-fibration over a surface $B_2$. 
What is important for us is that the $\mathbb P^1$-fibration  of the  base $B_3$ admits two sections, $S_+$ and $S_-$. Prior to taking into account the effect of the heterotic gauge bundle, these are the surfaces wrapped by 7-branes hosting the two $E_8$ factors of the dual heterotic string. 
  
  Consider first a situation where $E_8^{(2)}$ is unbroken by the gauge background.
  In the dual F-theory, the gauge degrees of freedom of this $E_8$ are localised on a stack of general $(p,q)$ 7-branes along the divisor $S_-$.
  A heterotic 5-brane instanton embeddeded into $E_8^{(2)}$ corresponds to a Euclidean D3-brane instanton wrapped on the divisor $S_-$. In counting the zero-modes of this instanton, one must take into account that the instanton is coincident to the non-abelian gauge stack.\footnote{For a D3-brane instanton not coincident to a 7-brane
  the zero mode counting proceeds most easily by the uplift to M-theory in terms of the cohomology groups $H^i(\pi^\ast(S_-),{\cal O})$, with $\pi^\ast(S_-)$
the divisor of the Calabi-Yau 4-fold obtained by restricting the elliptic fibration to $S_-$ \cite{Witten:1996bn, Blumenhagen:2010ja}. For D3-brane instantons along smooth divisors dual to heterotic NS5-brane instantons, this has been discussed in \cite{Palti:2020qlc}, exemplifying the contribution of such instantons to the superpotential. In the present case, however, this formula cannot be directly applied because $\pi^\ast(S_-)$ is singular as a result of the 7-branes along $S_-$.}
 We propose to count the zero modes on the instanton in the same way as for a spacetime-filling 7-brane \cite{Donagi:2008ca, Beasley:2008dc, Weigand:2018rez}:
 In absence of instanton flux, this gives the usual universal zero modes $(x^\mu, \theta^\alpha)$ (counted by $h^0(S_-, {\cal O}) =1$)
  and $\bar\tau^{\dot\alpha}$ (counted by $h^2(S_-, {\cal O} \otimes K_{S_-}) =1$, with $K_{S_-}$ the canonical bundle on $S_-$ )\footnote{Note that since $S_-$ hosts the second $E_8$ gauge group factor, it is not intersected by the residual components of the discriminant, which would be responsible, for instantons on divisors not wrapped by 7-branes, for the lifting of the $\bar\tau^{\dot\alpha}$ modes.}, as well as extra chiral and anti-chiral zero modes,  counted  by $h^1(S_-, {\cal O}) + h^2(S_-, {\cal O})$.
   For $S_-$ a section of $B_3$, the latter dimensions of cohomology groups vanish.
   What makes the generation of a superpotential possible is that the anti-chiral zero modes $\bar\tau^{\dot\alpha}$ are lifted by interactions with additional zero modes in the sector between the D3-branes and the 7-branes by a mechanism detailed in \cite{Akerblom:2006hx,Petersson:2007sc}. 
   
   For unbroken $E_8^{(2)}$, the D3-brane instanton along $S_-$ therefore acts as an $E_8$ gauge instanton contributing to the superpotential.
 As $E_8^{(2)}$ is broken
by an abelian gauge background to $\prod_m U(1)_m$, the general expectation is that this instanton continues to exist and gives rise to multiple individual instantons.
This picture can be confirmed most easily
for heterotic gauge bundles which map,
in the dual F-theory,
to a collection of line bundles on the 7-brane stack along $S_-$. 
 In particular, this is the case for the line bundles 
 pulled back from the base of the heterotic elliptic fibration $X$ which, as explained in Appendix \ref{App:3gen}, are tailor-made to ensure the vanishing of the chiral index for the charged matter fields from $E_8^{(2)}$.
   By a suitable choice of instanton flux, one can always find a D3-brane instanton that carries vanishing relative flux with respect to one of the effective 7-branes associated with each of the $U(1)_m$ gauge group factors. 
    Importantly, since the lifting of the 
    anti-chiral zero-modes $\bar \tau^{\dot \alpha}$ is still in place \cite{Petersson:2007sc}, this instanton contributes to the superpotential, but in general only with a factor involving products of vector-like charged matter fields; the latter absorb vector-like charged instanton zero modes \cite{Blumenhagen:2006xt,Ibanez:2006da}, collectively denoted as $(\lambda_j, \tilde \lambda_j)$, from the zero modes of strings between the instanton and the other 7-branes along $S_-$. Here it is important to note that these modes are counted by the same expression that counts the $U(1)_m$ charged fermions in spacetime, because the instanton wraps the same cycle as the 7-brane and carries the same gauge background. 
    For the  $U(1)_m$ instantons with $m\in {\cal I}$, for which no chiral charged fermions can exist, this guarantees that the charged instanton zero modes only come as vector-like pairs $(\lambda_j, \tilde \lambda_j)$ with a vanishing net chiral index. 
    The impact of the resulting monomials $(\phi_j \tilde \phi_j)^{n_{m,j}}$ in the superpotential (\ref{eq:superpotential_contribution_axion_potential}) 
   will be discussed in Section ~\ref{sec:light_fermions}.
 
These fluxed instantons are to be viewed as the analogues of gauge instantons for the abelian gauge group factors \cite{Petersson:2007sc}, in agreement with the fact that their exponential suppression factor is set by the inverse $U(1)_m$ coupling. 
 The requirement that the fluxed D3-brane instanton wraps the cycle $S_-$ rather than a multiple or fraction therefore means that the instanton
suppression factor is normalised such as to agree, at tree-level, with $\alpha^{-1}_{\rm YM,1}$ defined in (\ref{eq:YM_corrected_with_threshold-2}). 
In other words, the suppression factor corresponds to the inverse gauge coupling of the properly normalised abelian factor
 \begin{equation}
\widehat{U(1)}_m := \sqrt{\frac{4}{\eta_{i,i}}} U(1)_m \label{def-U(1)hat}\,,
 \end{equation}
with inverse gauge coupling
squared\begin{equation}
\widehat{\alpha}^{-1}_{{U(1)}_m} =  \alpha^{-1}_{\rm YM,1} + \widehat{\Delta}_i\,,
\end{equation}
in terms of the normalised non-universal abelian thresholds (\ref{def-hatdeltai}).
 Including the  latter, 
 the instanton suppression factors of each ${U(1)}_m$ factor begin to differ at subleading level. However, the instanton action associated with the specific linear combination $U(1)_D$ defined in (\ref{def-U1D}) is independent of these non-universal thresholds. In particular, this instanton is exponentially suppressed by 
 $\alpha_{\rm YM,1}^{-1}$ given by (\ref{def-U1D}) and couples to the linear combination $\theta_2$ defined in (\ref{eq:linear_comb_theta_12}). 

The discussion above implies that to leading order, the action of the described small gauge instantons from fluxed 5-branes can be estimated as $S_{\rm UV}={2\pi}/{\alpha_{\rm GUT}}$. 
By doing so, we used that the gauge couplings of both $E_8$ factors are the equal at tree level and assumed that the first $E_8$ coupling is comparable to the standard unification prediction, cf. 
(\ref{alphac-def}) and (\ref{alphaYM1}). This also implies that the leading order estimate for $g_{a\gamma}/m_a$ that we obtain from the UV instanton potential coincides for $\theta_2$ and $\varphi$. The non-universal 1-loop threshold corrections 
will be discussed in detail in Section \ref{sec_thresholds}.
In particular, these thresholds will distinguish between the potential for $\theta_2$ and $\varphi$.

 Disregarding these threshold corrections for the moment, we obtain a leading-order estimate for the axion potential in the EFT derived from the high-scale superpotential (see (\ref{eq:superpotential_contribution_axion_potential}))~\cite{Svrcek:2006yi,Conlon:2006tq}    
\begin{equation}\label{eq:ALP_potential_low-scale_susy}
  V(\theta_2,\varphi)= \kappa\frac{16\pi}{\alpha_{\rm GUT}}m_{\text{susy}}M_{\rm UV}^3e^{-2\pi/\alpha_{\rm GUT}}\cos (\theta_2)  + (\theta_2 \leftrightarrow \varphi)\,,
   \end{equation}
where $M_{\rm UV}\sim M_{\rm GUT}$ is a UV scale close to the cutoff of the 4d EFT.\footnote{In practice the UV scale is slightly above the cut-off of the 4d theory, $M_{\rm UV}^3=M_s^2M_P$, indicating that our estimate is conservative. See~\cite{Demirtas:2021gsq} for a recent discussion.} The quantity $\kappa$ is related to the one-loop determinant and contains additional chiral suppression beyond MSSM particles. (MSSM-only fermion content implies $\kappa\rightarrow 1$, since no chiral fermion comes from the second $E_8$). 
We will analyse these effects in Section \ref{sec:light_fermions}, where we will find that setting $\kappa =1$ 
 offers a conservative estimate and serves as a reliable lower-bound on the ALP potential.

The leading order estimates that are derived from the potential $V(\theta_2,\varphi)$ are shown in  Figure~\ref{fig:ParameterSpace} for different values of $m_{\rm susy}$ and $\alpha_{\rm GUT}$. In all three cases, the SM-charged matter corresponds to 3-generation MSSM-like constructions with no exotics. The large enhancement of $g_{a\gamma}/m_a$ for the ALP in cases with high-scale supersymmetry (gold dashed line) is due to the decreasing value of the unified gauge coupling, $\alpha_{\rm GUT}$. Any new physics between the EW and KK scale will result in this predicted line moving to the right in the $(g_{a\gamma},m_a)$ plane. 

\subsection{UV instanton potential in the presence of light fermions}\label{sec:light_fermions}
In this section we consider the effect of 
$n_f = \sum_j n_{m,j}$ vector-like pairs of light fermions
with $U(1)_m$ charge on the axion potential in Eq.~\eqref{eq:ALP_potential_low-scale_susy}. 
 Light pairs of vector-like fermions can arise 
from fermions that are massless prior to supersymmetry breaking and subsequently acquire a mass. In the UV, these fermions are treated as massless and appear as the monomial $\prod_j (\phi_j \tilde \phi_j)^{n_{m,j}}$ in the superpotential (\ref{eq:superpotential_contribution_axion_potential}). 
Importantly, the same supersymmetry breaking effect that lifts these modes in the 4d EFT also induces a mass of the same scale for the NS5-brane instanton zero modes $(\lambda_j, \tilde \lambda_j)$ whose absorption in the instanton path integral induced the monomials in (\ref{eq:superpotential_contribution_axion_potential}). 
 This is clearest again in the F-theory dual picture, where the NS5-brane instanton is described by a D3-brane instanton parallel to one of the 7-branes. The instanton then couples to the string theory background in the same way as the spacetime-filling 7-brane on which the charged modes of the 4d EFT are located.

The net effect of these mass terms is, first, 
to replace the monomial $\prod_j (\phi_j \tilde \phi_j)^{n_{m,j}}$ in (\ref{eq:superpotential_contribution_axion_potential}) by a chiral suppression term  $(m_f/M_{\rm UV})^{n_f}$. 
 To arrive at this conclusion via the rules of string instanton calculus, one takes into account a small mass term of order $m_f$ for the charged modes $(\lambda_j, \tilde \lambda_j)$ in the instanton effective action that appears in the exponent and saturates the zero modes in the path integral by bringing down a suitable product of mass terms. This is the stringy version of the general logic of gauge instanton calculus in field theory, where light fermions (but with non-zero mass) in the 4d EFT induced by the instanton background are absorbed by mass insertions or Yukawa interactions with scalar fields.  

When the fermion masses are smaller than the UV scale at which the instantons contribute to the amplitude, $m_f\ll M_{\rm UV}$, the chiral suppression factor can be very small\footnote{If instead the light fermions are coupled to scalars the suppression factor is given by the product of Yukawa couplings, $\prod_i^{n_f}\frac{y_i}{4\pi}$, since scalar loops can saturate the fermionic zero-modes. For example, in the case of the QCD axion this has been argued to make UV instantons relevant and to raise its mass
with respect to the IR contribution~\cite{Holdom:1982ex,Flynn:1987rs,Rubakov:1997vp,Choi:1998ep,Agrawal:2017ksf,Gaillard:2018xgk,Fuentes-Martin:2019bue,Csaki:2019vte}.}
\begin{equation}\label{eq:chiral_supp}
    \kappa = (m_f/M_{\rm UV})^{n_f}\ll 1\,.
\end{equation}
As one can see in Eq.~\eqref{eq:superpotential_contribution_axion_potential}, the axion potential receives multiple contributions coming from each of the $U(1)$ symmetries that are unbroken by the gauge background. This suggests that in order to suppress the axion potential via chiral suppression, there should be fermions charged under every single $U(1)_m$. For the sake of simplicity, and to be as conservative as possible, we focus only on the contribution from the linear combination of $U(1)_m$ factors that contributes to hypercharge, neglecting instantons coming from \textit{purely dark} $U(1)$s. In this case, light fermions have hypercharge and we know, from phenomenological constraints, that their mass is between the EW scale and the compactification scale, $v_{\rm EW}\lesssim m_f \lesssim M_{\rm UV}$. We remark, however, that if one of the $U(1)_m$ has no light charged fermion, the corresponding instanton will likely dominate the axion potential.

The presence of light fermions 
 has another effect in addition to inducing a chiral suppression factor (\ref{eq:chiral_supp}):
 it modifies the running of the $U(1)_{Y}$ gauge coupling by changing the beta function at energies above $m_f$. 
Since the hypercharge coupling is well measured at around the EW scale we can use the beta function to relate an axion potential generated by a UV instanton with MSSM matter only, where the gauge coupling is $\alpha_{\rm GUT}\sim 1/24$, to the case with $n_f$ light fermions in addition to the MSSM. To illustrate the effect of the light fermions, we will assume the simplest situation where  the mass of the chiral superfields follows the relation $m_{\rm EW}\lesssim m_{\rm susy}\lesssim m_f$, although the net result does not rely on this concrete ordering.

Taking into account the light fermions, the value of the gauge coupling at the compactification scale can be written as 
\begin{equation}\label{eq:UV_gauge_coupling_with_light_fermions}
    \alpha_Y^{-1}(M_{\rm UV}) = \alpha_Y^{-1}(m_f) + \frac{\beta_Y}{2\pi}\ln (M_{\rm UV}/m_f) = \frac{5}{3}\alpha^{-1}_{\rm GUT} + \frac{\Delta \beta_{Y}}{2\pi}\ln (M_{\rm UV}/m_f) \,.
\end{equation}
Here we have used the fact that the 1-loop beta function, $\beta_Y$, contains the \textit{known} MSSM contribution and the correction induced by the new, light superfields, $\beta_Y=\beta_Y^{\rm MSSM}+\Delta \beta_{Y}$. Assuming, for concreteness, $k_1^{(2)}=1$ and fermions of mass $m_f\gtrsim m_{\rm susy}$ and charge $q_f=1$,  we have $\Delta\beta_{Y}=-2n_f$. 

Recall from Section \ref{sec:charged_matter_intermedite_scales} that the correction induced by light fermions charged only under hypercharge tends to increase the deviation from gauge coupling unification
in models with non-standard hypercharge embedding.
 Indeed, as the light fermions are charged under the unbroken $U(1)$s from $E_8^{(2)}$, the change to the gauge coupling by the new states is not GUT symmetric. Only $U(1)_Y$ receives contributions from the second $E_8$ and, therefore, similar corrections to $\beta_{SU(2)}$ and $\beta_{SU(3)}$ do not arise; importantly, the sign of the correction is opposite to what would be needed to compensate for the effect of $k_1^{(2)}$ in (\ref{eq:hypercharge_coupling_with_threshold}).
 The correction from charged $SU(3)_C$ and $SU(2)_w$ singlets must therefore be small.


In any case, defining the non-GUT symmetric gauge coupling shift as
\begin{equation}
    \Delta\alpha^{-1}\equiv -\frac{n_f}{\pi}\ln (M_{\rm UV}/m_f)\,,
\end{equation}
we can write the chiral suppression factor (\ref{eq:chiral_supp}) as 
\begin{equation} \label{kappa-2}
\kappa = e^{\pi\Delta\alpha^{-1}} \,.
\end{equation}
This relation indicates that the chiral suppression  becomes less relevant, $\kappa \rightarrow 1$, as the non-GUT symmetric correction $\Delta \alpha^{-1}$ is made small (note this quantity is negative due to the $U(1)$ beta function).

Apart from leading to the identification of the chiral suppression factor  (\ref{eq:chiral_supp}) as  (\ref{kappa-2}), once the IR gauge couplings are fixed to match the measured values, the effect induced by the light charged fermions on the gauge coupling 
 reduces the UV instanton action, $S_{\rm UV}$, and the ALP shift symmetry breaking becomes more relevant than without the new fermions. The simplest way to see this is to write the instanton action, $S_{\rm UV}\sim \frac{2\pi}{\alpha(M_{\rm UV})}$, in terms of Eq.~\eqref{eq:UV_gauge_coupling_with_light_fermions}. After some algebra, and using the expression for the chiral suppression factor in Eq.~\eqref{eq:chiral_supp}, one can obtain the axion potential from 5-brane instantons including the effect of $n_f$ light fermions with $U(1)$ charge $q_f=1$,
\begin{equation}\label{eq:potential_with_light_fermions}
    V(\theta_2,\varphi) = \frac{16\pi}{\alpha_{\rm GUT}}  \left ( \frac{M_{\rm UV}}{m_f} \right )^{n_f} m_{\rm susy}M_{\rm UV}^3 e^{-2\pi/\alpha_{\rm GUT}} 
\cos (\theta_2)  + (\theta_2 \leftrightarrow \varphi)  \,.
\end{equation}
We find that the axion mass from UV gauge instantons in a theory with light, charged fermions grows as~$\sim \left ( \frac{M_{\rm UV}}{m_f} \right )^{n_f/2}$ with respect to a theory with MSSM matter only, see~\eqref{eq:ALP_potential_low-scale_susy}. Note that this formula is only valid for $m_f \gtrsim m_{\rm susy}\gtrsim m_{\rm EW}$ since it was derived using Eq.~\eqref{eq:UV_gauge_coupling_with_light_fermions} but similar relations follow for different orderings of the mass scales.

Summarizing, the modified axion potential in Eq.~\eqref{eq:potential_with_light_fermions} suggests that the estimate for the axion potential induced by UV instantons with MSSM matter in Eq.~\eqref{eq:ALP_potential_low-scale_susy}  is a reliable, conservative estimate. Once the IR gauge couplings are fixed to the observed values, any additional light degree of freedom charged under the gauge group $\prod_m U(1)_m$, be it a fermion or a scalar, will only make the effects of UV instantons more sizeable, and therefore the coupling-to-mass ratio $g_{a\gamma}/m_a$ for $\theta_2$ and $\varphi$ smaller. 
In Figure \ref{fig:ParameterSpace} we consider some benchmark points for the SUSY scale and the UV value of the gauge coupling, including $m_{\rm susy}= 10^4$ GeV , $\alpha_{\rm GUT} = 1/24$ and 
$m_{\rm susy}= 10^9$ GeV , $\alpha_{\rm GUT} = 1/30$. 
For completeness we also include the prediction for $g_{a\gamma}/m_a$ in a case with no low-scale supersymmetry where $\alpha_{1}\sim 1/37$ at around the GUT scale.\footnote{We have assumed that the CY has a volume $V_{\rm CY}\sim r^6$, with $r^{-1}\sim M_{\rm GUT}= 2\times 10^{16}$ GeV to avoid excessively fast proton decay.}

\subsection{Worldsheet instantons}\label{sec:worldsheet_and_ALPs}
In this section we comment on additional instanton contributions to the masses of the model dependent axions, $b_i$, and the 5-brane axions, $\tilde b_r$, which have no field theory analog in terms of small gauge instantons.

Consider first the model dependent axions $b_i$. These can obtain a mass from heterotic worldsheet instantons~\cite{Wen:1985jz} due to Euclidean fundamental heterotic strings wrapping suitable 2-cycles.
  In order for these instantons to contribute to the axion masses via a potential, additional instanton zero modes, if present, must be absorbed and summation of the instantons must not lead to a cancellation. This may not always be possible depending on the details of the geometry \cite{Beasley:2003fx}.
  {\it If} an axion mass is generated, it is expressed exponentially by the worldsheet instanton action ${S_{\rm ws}=2\pi \text{Vol}(C_\alpha)}$, where the volume of the 2-cycle is given by the integral of the K\"ahler form,
\begin{equation}\label{eq:2-cycle_vol}
    \text{Vol}(C_\alpha) = \int_{C_\alpha} J = t^i \int_{C_\alpha} {D_i} = t^i Q_{i\alpha}\,.
\end{equation}
 Here $D_i \in H^2(X,\mathbb Z)$ is a basis of the K\"ahler cone,
$t^i\geq 0$ are the K\"ahler parameters and
$Q_{i\alpha} = \int_{C_\alpha} {D_i} \in \mathbb{Z}$.

 As we will now discuss, the worldsheet instanton action, $S_{\rm ws}$, is generically expected to be smaller than the total volume of the Calabi-Yau.
  The latter is given in terms of the K\"ahler parameters and the triple 
 intersection numbers
 \begin{equation}
    \kappa_{ijk} = \int_X {D_i \wedge D_j \wedge D_k}
\end{equation}
as
\begin{equation}\label{eq:CY_volume}
    \mathcal{V} = \frac{1}{6} \kappa_{ijk} \, t^i t^j t^k\,,
\end{equation}
and is related to the gauge coupling as 
\begin{equation} \label{alphaGUT-a}
\frac{1}{\alpha_{\rm GUT}}=\frac{\mathcal{V}}{g_s^2}\,.
\end{equation}
As discussed above, this fixes the UV instanton action $S_{\rm UV}$ (NS5-branes wrapping the entire compact space), and furthermore shows the well-known fact that in perturbative heterotic models, ${\cal V} \lesssim 1/\alpha_{\rm GUT}$.

To make more robust statements about the masses of model dependent axions $b_i$,
the naive expectation $\text{Vol}(C_\alpha) \lesssim {\cal V}$ (and therefore $  \text{Vol}(C_\alpha)\lesssim 1/\alpha_{\rm GUT}$ in peturbative settings) must be put on more solid grounds. 
 A crucial role is played by the fact that in the K\"ahler cone basis, in which $t^i\geq 0$, the triple intersection numbers satisfy
$\kappa_{ijk} \geq 0$.
 It is therefore clear that those moduli with $\kappa_{iii} > 0$ must obey $t^i \lesssim {\cal V}$.
In particular, if $\kappa_{iii} > 0$ for a large number of moduli, these are even subject to
 $t^i \ll {\cal V}$.

At the level of classical K\"ahler geometry, a potential loophole to this conclusion arises for moduli with $\kappa_{iii} =0$,\footnote{Recall that by Oguiso's theorem, in this case, since $D_i$ is a K\"ahler basis, the Calabi-Yau threefold $X$ must admit either a torus- or a K3/$T^4$ fibration.} which at first sight can
 obey $t^i\gg 1$ (without compromising  ${\cal V} \lesssim 1/\alpha_{\rm GUT}$) if
their contribution to ${\cal V}$ appears multiplied by small K\"ahler parameters $t_j,t_k\ll 1$, so that $t_it_jt_k\sim O(1)$.  
 However, this situation is at odds with keeping control of the $\alpha^\prime$ expansion, in the following sense:
  To obtain parametrically large $t^i\gg 1$ in this way without increasing the volume ${\cal V}$ of the Calabi-Yau $X$, the latter must admit an elliptic or a K3/$T^4$-fibration (see \cite{Lee:2019wij} in a different context) with the base becoming large while the fiber shrinks such that large 
 2-cycles receive a contribution from the large base. Worldsheet instantons along the fiber would then be unsuppressed and significantly modify the classical geometry at the quantum level.
  For example, for K3/$T^4$ fibrations in Type IIA compactifications \cite{Lee:2019wij} (with $\mathcal{N}=2$ supersymmetry) this effect  obstructs the limit of vanishing fiber volume in the quantum moduli space by bounding the fiber moduli at ${\cal O}(1)$ in string units. Similar behaviour is expected for heterotic compactifications, which preserve only $\mathcal{N}=1$ supersymmetry at the compactification scale, so that no parametric suppression of worldsheet instantons from large 2-cycle volumes can occur at fixed volume ${\cal V} \lesssim 1/\alpha_{\rm GUT}$.
  
Obtaining large 2-cycles in string units hence necessitates correspondingly large ${\cal V}$. In view of (\ref{alphaGUT-a}) this requires
 co-scaling $g_s^2$ as ${\cal V}$ becomes large. It is precisely in this limit that the gauge threshold corrections become $O(1)$. Their effect on the UV instanton action will be studied in Section \ref{sec_thresholds}. In any event, the limit $g_s^2\gg 1$  leaves the regime of perturbative heterotic string theory, which is the framework under consideration in this work.

To summarize, in perturbative (in $g_s$) scenarios and especially with $h_{11}\gg 1$ some, if not all, of the 2-cycle volumes will be not much larger than string scale, leading to worldsheet instanton actions $S_{\rm ws}\sim O(1-10)$. See also~\cite{Leedom:2025mlr} for a recent study of the impact of worldsheet instantons on model dependent axions reaching similar conclusions.

As an extreme case consider a scenario where 
all $t^i\sim O(1)$, as required in particular for large $h_{11}$.
In this situation one has $S_{ws}^i\sim O(1)\ll S_{\rm UV}$.
 Modulo the important assumption that these worldsheet instantons indeed contribute to the superpotential, they induce a large mass for the model dependent axions $b_i$, which become heavy and can be integrated out.
 In models without NS5-brane axions, it is possible to show that, as a result, the two special linear combinations 
$\theta_1$ and $\theta_2$ of axions defined in (\ref{eq:linear_comb_theta_12}) 
 coincide after integrating out all the heavy model dependent axions $b_i$ and complex scalar phases $c_j$~\footnote{Only phases $c_j$ that also mix with the model independent axion $a$ survive the effect of worldsheet instantons. However, due to this mixing, they couple universally to all the gauge bosons.},
%
\begin{align}
    \theta_1|_{\tilde b_r =0} = a +\sum_i n_i b_i + \sum_j m^{(1)}_j c_j  \rightarrow a  \,,\\
    \theta_2 |_{\tilde b_r =0} = a -\sum_i n_i b_i + \sum_j m^{(2)}_j c_j \rightarrow a \,.
\end{align}
Since $\theta_1=\theta_2=a$, the only light axion coupled to gauge boson is the QCD axion (due to its coupling to the first $E_8$) and the bound \eqref{eq:coupling_mass_relation} is automatically satisfied even for a non-standard embedding of hypercharge. Similarly, the non-universal axion $\varphi$ (see (\ref{eq:varphi_coupling_to_gaugebosons2})) becomes heavy and is integrated out. Heavier model dependent axions $b_i$ couple with unsuppressed couplings to photons, but they also satisfy Eq.~\eqref{eq:coupling_mass_relation} due to their large mass.

Let us now turn to 
the NS5-brane axions, $\tilde b_r$.
They can receive an analogous contribution to their mass from non-perturbative string instantons due to Euclidean non-critical strings wrapped on curves. Recall that from the perspective of heterotic M-theory, the non-critical strings arise from M2-branes suspended along the M-theory interval between M5-branes or between one M5-brane and one of the $E_8$ planes, while M2-branes suspended between both $E_8$-planes map to the fundamental heterotic string.   If the M5-branes are located at generic points of the M-theory interval, this leads to an additional order one suppression of the instanton action compared to perturbative worldsheet instantons, corresponding to the fraction of the M-theory interval wrapped by the suspended M2-brane that maps to the Euclidean non-critical string. In such generic situations, the instanton generated masses of the NS5-brane axions $\tilde b_r$ are hence enhanced compared to the those of the axions $b_i$. 
  
To summarize, while it is hard to make precise predictions for the mass of the axions $b_i$ and $\tilde b_r$ that hold in full generality, 
 moderately small 2-cycles 
are typically associated with heavy model dependent axions coupled to photons with interesting implications in cosmology. 
As a concrete example let us take a cycle with $\text{Vol}(C)=10$. In this case, for an axion coupled via the GS counter-term, one obtains $g_{a\gamma}/m_a=4\times 10^{-13} \text{ GeV}^{-2}$ corresponding to the orange line in Figure~\ref{fig:ParameterSpace}. Model dependent axions are expected to have large decay constant, close to the GUT scale, $f_{b_i}\sim 10^{16}$ GeV~\cite{Svrcek:2006yi}. If such a heavy model dependent axion is produced in the early universe via misalignment, freeze-in or similar mechanisms, we find that X-rays, CMB, BBN, and other cosmological observables may offer stringent constraints. One concrete example, as shown recently in~\cite{Yin:2025amn} in the context of type IIB strings, is that the decay of heavy axions produced via freeze-in can be constrained with reionization. In the case of GUT symmetric couplings, efficient production in supernovae and subsequent decay into photons offers strong constraints~\cite{Benabou:2024jlj}, see dotted red line in  Figure~\ref{fig:ParameterSpace}. These examples illustrate how heavy model dependent axions can have a non-trivial impact in early universe cosmology due to their photon coupling. We leave a more quantitative study of their implications for future work.

\subsubsection{The small string coupling limit $g_s^2\ll 1$ }\label{sec:small_gs}
In the limit that the string coupling becomes weak, some effects including threshold corrections become parametrically small. This occurs because being induced by heavy stringy modes,  they are suppressed by $g_s^2$. However, in the heterotic string, the total volume and the string coupling are related via the gauge coupling, see Eq.~\eqref{alphaGUT-a}. For this reason, the string coupling is bounded by $g_s^2\gtrsim \alpha_{\rm GUT}$. As the string coupling decreases and reaches $g_s^2\sim \alpha_{\rm GUT}$, the total volume of the compact space reaches $\mathcal{V}\rightarrow 1$ and we enter a non-geometric regime. At this point the typical size of two-cycles becomes smaller than 1 in string units and the $\alpha^\prime$ expansion breaks down.



This fact has important implications for  axion physics. We expect the model independent axion $a$ to be unaffected  -- its mass comes exclusively from either IR gauge instantons of confining interactions or from NS5-branes, which have an action $S_{\rm UV}\sim 2\pi/\alpha_{\rm GUT}$  insensitive to the aforementioned string coupling. However, things are different for the model dependent axions $b_i$, whose shift-symmetry is broken by worldsheet instantons. As one approaches the weak coupling limit, the typical worldsheet action decreases as $g_s^2$ with respect to the NS5-brane action which remains fixed. For this reason, similar to situations where $t^i\rightarrow 1$ due to large $h_{11}$, for small $g_s$ one generically expects that the model dependent axions $b_i$ decouple from the low-energy phenomenology. 
 It is then clear already from this argument that the bound in Eq.~\eqref{eq:coupling_mass_relation} holds even for a non-standard embedding of hypercharge.

\section{Realistic models with non-standard hypercharge embedding and $g_{a\gamma}/m_a$} \label{sec_thresholds}
So far we have described the relevant ingredients to estimate the coupling-to-mass ratio for ALPs in heterotic string models with non-standard embedding of hypercharge. We have found that there are two different ALPs, $\theta_2$ and $\varphi$, that couple to photons and define new lines in the parameter space with constant ratios $g_{\theta_2\gamma}/m_{\theta_2}$ and $g_{\varphi\gamma}/m_{\varphi}$, which we estimated for different models in the absence of threshold corrections, $|\Delta_Y| \ll 1$, or new charged matter beyond the SM representations, see  Figure~\ref{fig:ParameterSpace}. In the case $\Delta_i\ll 1$ all small gauge instanton actions are well approximated by $S\approx \frac{2\pi}{\alpha_{\rm YM}}$ and we expect $g_{\theta_2\gamma}/m_{\theta_2}\sim g_{\varphi\gamma}/m_{\varphi}$.

These minimal scenarios, however, are not realistic due to the tree-level hypercharge gauge coupling prediction at the GUT scale,
\begin{equation}\label{eq:matching_sec.4}
    \frac{1}{\alpha_Y}  = \frac{5/3}{\alpha_{\rm GUT}}+ \frac{k_1^{(2)}}{\alpha_{\rm GUT}} \,,
\end{equation}
which does not reproduce the measured gauge couplings at low energies unless $k_1^{(2)} \ll 1$. Indeed, in Figure \ref{fig:ParameterSpace} we have assumed for demonstrative purposes that $k_1^{(2)} =1$, as is the case e.g. in the models presented in Section \ref{sec:Timo's_model}.
As discussed in Section \ref{sec:unif_gauge_coupling}, the IR value of the gauge coupling can be corrected in different ways, including stringy threshold corrections, light charged matter or a small level of embedding $k_1^{(2)}$. In this section we will estimate how these different mechanisms affect the leading order predictions for the coupling-to-mass ratio shown in  Figure~\ref{fig:ParameterSpace}.
Due to their model dependence, and to understand qualitatively the implications of each possibility, we study these cases separately. We note, however, that in realistic models they could be present simultaneously. 

We separate these models in two large classes depending on the contribution from threshold corrections, $\Delta_Y$ (see Eq.~\eqref{eq:threshold_hypercharge}). First we consider, in Section \ref{sec_smallthesholds}, models where threshold corrections are small. These include models with charged matter at intermediate scales that changes the running of the gauge couplings and scenarios with small level of embedding in the second $E_8$, $k_1^{(2)}$. In Section \ref{sec_theorieswithlargethresholds}, we consider large threshold corrections from stringy states correcting the gauge couplings. Assuming that the hypercharge correction is $\Delta_Y<0$ so that \eqref{eq:matching_sec.4} is corrected, we will split the discussion into theories with positive and negative non-universal abelian corrections, $\widehat{\Delta}_i$.

\subsection{Theories with small threshold corrections, $\Delta_Y\rightarrow 0$} \label{sec_smallthesholds}

One way to construct theories with small one-loop threshold corrections is by considering a region in K\"ahler moduli space where the specific linear combination of curve volumes entering the expression for the thresholds is small.
 Alternatively, the threshold corrections are small if $g_s^2\ll 1$ with the GUT gauge coupling $\frac{1}{\alpha_{\rm GUT}}=\frac{\mathcal{V}}{g_s^2}$ kept fixed. This reduces the total volume $\mathcal{V}$ as well as the 2-cycle volumes, see Eq.~\eqref{eq:2-cycle_vol}. For this reason, as anticipated in Section \ref{sec:small_gs}, in both scenarios we expect that the worldsheet instanton action is also reduced, $S_{\rm ws\, ,\,\alpha}\propto \int_{C_\alpha}J$, and the typical mass of model dependent axions will increase. 
In the most extreme case where all the worldsheet instanton actions are small, $S_{\rm ws\, ,\,\alpha}\sim O(1)$, every model dependent axion receives a large mass which in some cases could be close to the KK scale. 

In this section we assume that $\Delta_Y$ is small and study the effects on $g_{a\gamma}/m_a$ for two different scenarios that allow to obtain the measured IR gauge couplings.

\subsubsection{Models with charged matter at intermediate scales}\label{sec:intermediate_states_ALP_potential}
New matter with $SU(2)_w$ and $SU(3)_C$ charge at intermediate scales can help to achieve unification in some scenarios. {As described in Section \ref{sec:charged_matter_intermedite_scales}, however, in the simply connected line bundle models that we consider the new charged states cannot solve the issue of gauge coupling unification by themselves, but at best in combination with additional mechanisms such as threshold corrections or a small level of embedding, $k_1^{(2)}$. We cannot rule out, though, that in more complicated models light states might be fully responsible for the unification of couplings.}
See \cite{Font:1990uw} for an early discussion in heterotic orbifolds.

At any rate, when the IR gauge couplings are fixed to the measured values, it is unavoidable that the new states shift the value of the GUT gauge coupling,
\begin{equation}
\alpha_{\rm GUT}^{-1}=\alpha_{\rm initial}^{-1}+\Delta \alpha^{-1} \,.
\end{equation}
This has important implications for axions. Since the shift $\Delta\alpha^{-1}$, given in Eq.~\eqref{eq:shift_GUT_gauge_coupling}, is negative, the unified gauge coupling $\alpha_{\rm GUT}$ is stronger than the initial prediction, $\alpha_{\rm initial}$. 
Neglecting momentarily threshold corrections (which will be considered later), it can be easily seen that the action for the UV instanton is reduced,
\begin{equation}
    S_{\rm UV}^{(i)}=\frac{2\pi}{\alpha_{\rm GUT}}<\frac{2\pi}{\alpha_{\rm initial}}\,.
\end{equation}
This indicates that the shift symmetry breaking effects for $\theta_2$ and $\varphi$ are enhanced with respect to the original estimates in  Figure~\ref{fig:ParameterSpace}.
We also recall that the minimal contribution from $U(1)$ instantons arises when there are no light fermions from gauginos in  $E_8^{(2)}$, meaning that there is no additional chiral suppression. 
Such chiral suppression is only possible if the fermions have hypercharge, but this would require smaller $\alpha_{\rm GUT}^{-1}$  resulting in more sizeable shift-symmetry breaking effects, as we discussed around Eq.~\eqref{eq:potential_with_light_fermions}.

As an illustrative example of the significant impact of new states at intermediate scales, consider the following: It can be shown that in any scenario with intermediate states arriving at a unified gauge coupling $\alpha_{\rm GUT}\gtrsim 1/22$ the induced potential for $V(\theta_2,\varphi)$ then has a barrier height comparable to that of the QCD axion (assuming chiral suppression by SM matter only). This demonstrates that using exotic charged matter to obtain the observed IR gauge couplings has a big impact and decreases $g_{a\gamma}/m_a$ for $\theta_2$ and $\varphi$ with respect to the minimal estimates in  Figure~\ref{fig:ParameterSpace}.

\subsubsection{Models with small level of embedding $k_1^{(2)}$}
\begin{figure}[t]
    \centering
    \includegraphics[scale=0.4]{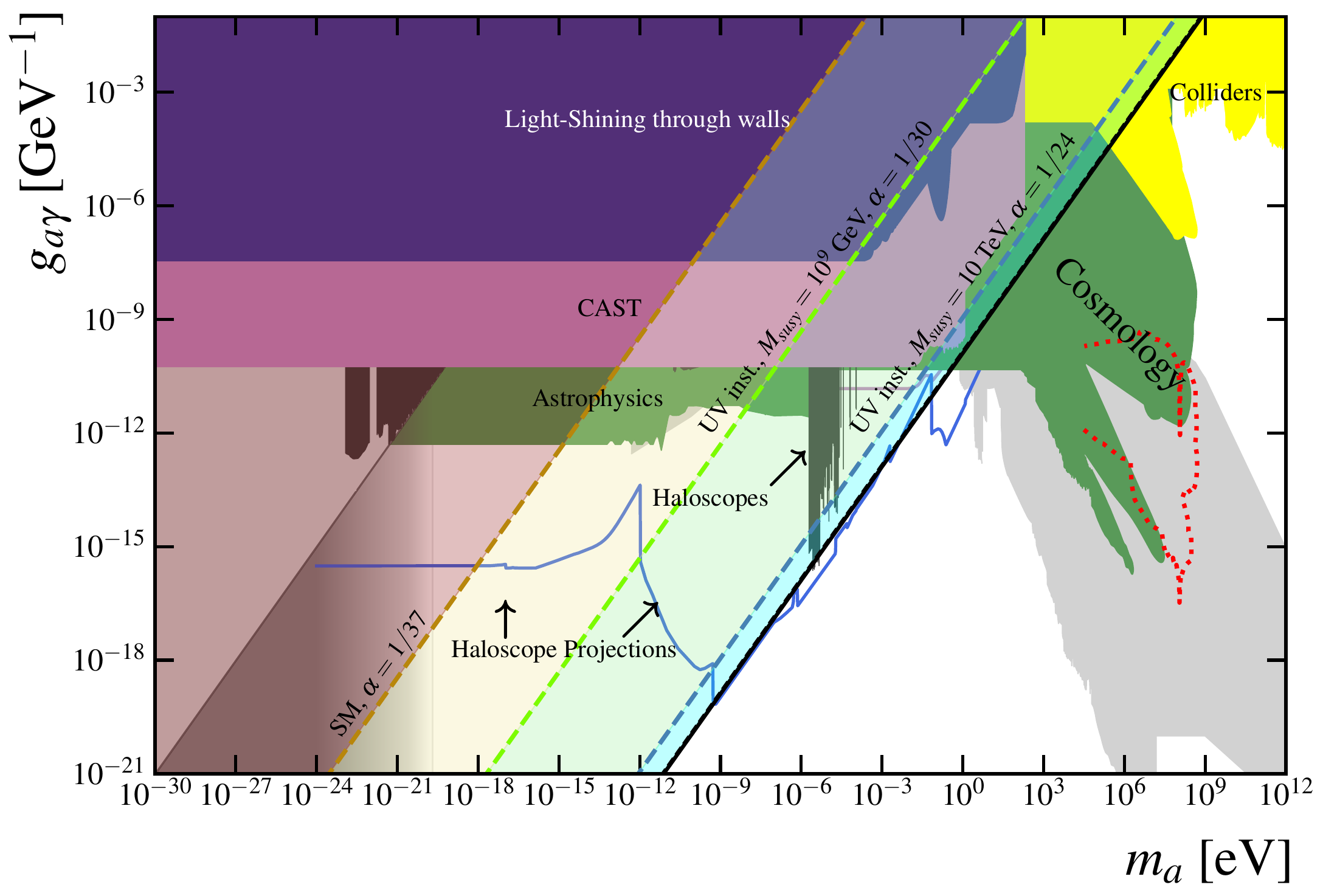}
    \caption{Same as  Figure~\ref{fig:ParameterSpace} but with small level of $U(1)_Y$ embedding into the second $E_8$, $k_1^{(2)}=1/9$. The smaller level of embedding leads to $E/N=41/16$ for $\theta_1$ and shifts the solid black line closer to the standard GUT prediction. The instanton action and axion masses are unchanged but the $\theta_2$ coupling to photons is reduced as $g_{\theta_2\gamma}\propto (k_1^{(2)})^2$. The coupling-to-mass ratio for $\varphi$, $g_{\varphi\gamma}/m_{\varphi}$ also vanishes in the limit of small $k_1^{(2)}$ (see text for details). For simplicity other lines with confining sectors are not shown.}
     \label{fig:ParameterSpace_with_small_k1}
\end{figure}
A different way to recover the measured gauge couplings at low energies without large threshold corrections is by engineering models with small level of embedding. 
This can occur in theories where the ratio $k_1^{(2)} = \frac{5}{3} \left(\frac{\kappa_{0,0} \ell_0}{\kappa_{1,1} \ell_1}\right)^2 \frac{\eta_{1,1}}{\eta_{0,0}}$ is small. Simple examples include theories where the line bundle in $E_8^{(2)}$ is rescaled by a moderately large value of $\ell_1 > 1$, as exemplified in Appendix \ref{app_ksmaller1} in a model with $\ell_1 = 3$. However, this coefficient cannot be arbitrarily enhanced because the first Chern classes are integrally quantised 
and the Bianchi identity restricts the maximal amount of gauge background. We therefore expect that in this way, at best values for $k_1^{(2)}$ of the order of $\frac{1}{10} - \frac{1}{100}$ are possible.

From Eq.~\ref{eq:matching_sec.4}, we find that in these models the prediction $\sin^2\theta_w=3/8$ at $M_{\rm GUT}$ is approximately recovered as $k_1^{(2)} \ll 1$.  
While having small $k_1^{(2)}$ does not modify the mass of $\theta_2$ -- the UV instanton action is unchanged -- the coupling to photons depends on this quantity. From Eq.~(\ref{eq:eff_Lag_non-standard_embedding}) we infer that the (dimensionful) coupling to photons scales quadratically with the level of embedding,
\begin{equation}\label{eq:g_a_in_small_k1_2}
    g_{\theta_2\gamma} = \frac{\alpha_{\rm em}}{2\pi}\frac{(k_1^{(2)})^2}{k_1^{(1)}+k_1^{(2)}}\frac{1}{f_{\theta_2}}  \,.
\end{equation}

This is just as expected since in the limit $k_1^{(2)}\rightarrow 0$ one should recover the standard GUT predictions where hypercharge is only embedded into a single $E_8$. Specifically, both the value of the gauge couplings and the weak mixing angle, as well as the axion couplings will follow the standard heterotic case~\cite{Agrawal:2024ejr} where only the QCD axion couples sizeably to photons and the bound \ref{eq:coupling_mass_relation} holds. In  Figure \ref{fig:ParameterSpace_with_small_k1} we show an example with $k_1^{(2)} = \frac{1}{9}$ (see Appendix~\ref{app_ksmaller1} for details about the model).

The coupling to hypercharge of the non-universal ALP $\varphi$ derived in \ref{sec_varphi} should also vanish as we recover the standard GUT prediction. This is slightly more subtle because it is not directly related to $k_1^{(2)}$ and is instead related to the fact that $U(1)_Y$ is a linear combination of line bundles. Let us parametrize the departure from the standard GUT prediction as
\begin{eqnarray}
    \alpha_Y^{-1}=5/3\alpha_{\rm GUT}^{-1}+\varepsilon\,,
\end{eqnarray}
where $\varepsilon=k_1^{(2)}\alpha_{\rm GUT}^{-1}+\Delta_Y$ is small. In the limit $k_1^{(2)}\rightarrow 0$, the only way to recover the GUT prediction is if $\Delta_Y$ vanishes at the same rate. As we discussed above, one way to make threshold corrections vanish is by taking a small $g_s$ limit. In this case, it is known that the model-dependent axion decay constant increases as $f_{b_i}^2\propto 1/g_s^2$~\cite{Svrcek:2006yi}. Therefore the coupling of $\varphi$ to gauge bosons decreases as $g_s$ decreases. Additionally, for a given model, as $g_s$ decreases so does the typical size of two-cycles and therefore the worldsheet instanton action. For this reason, we expect $g_{\varphi\gamma}/m_{\varphi}$ to decrease very fast with decreasing hypercharge threshold correction $\Delta_Y$. Indeed, using holomorphy, it can be seen that the coupling to photons of $\varphi$ vanishes as the GUT prediction is recovered.

\subsection{Theories with large threshold corrections} \label{sec_theorieswithlargethresholds}
\begin{figure}[t]
    \centering
    \includegraphics[scale=0.4]{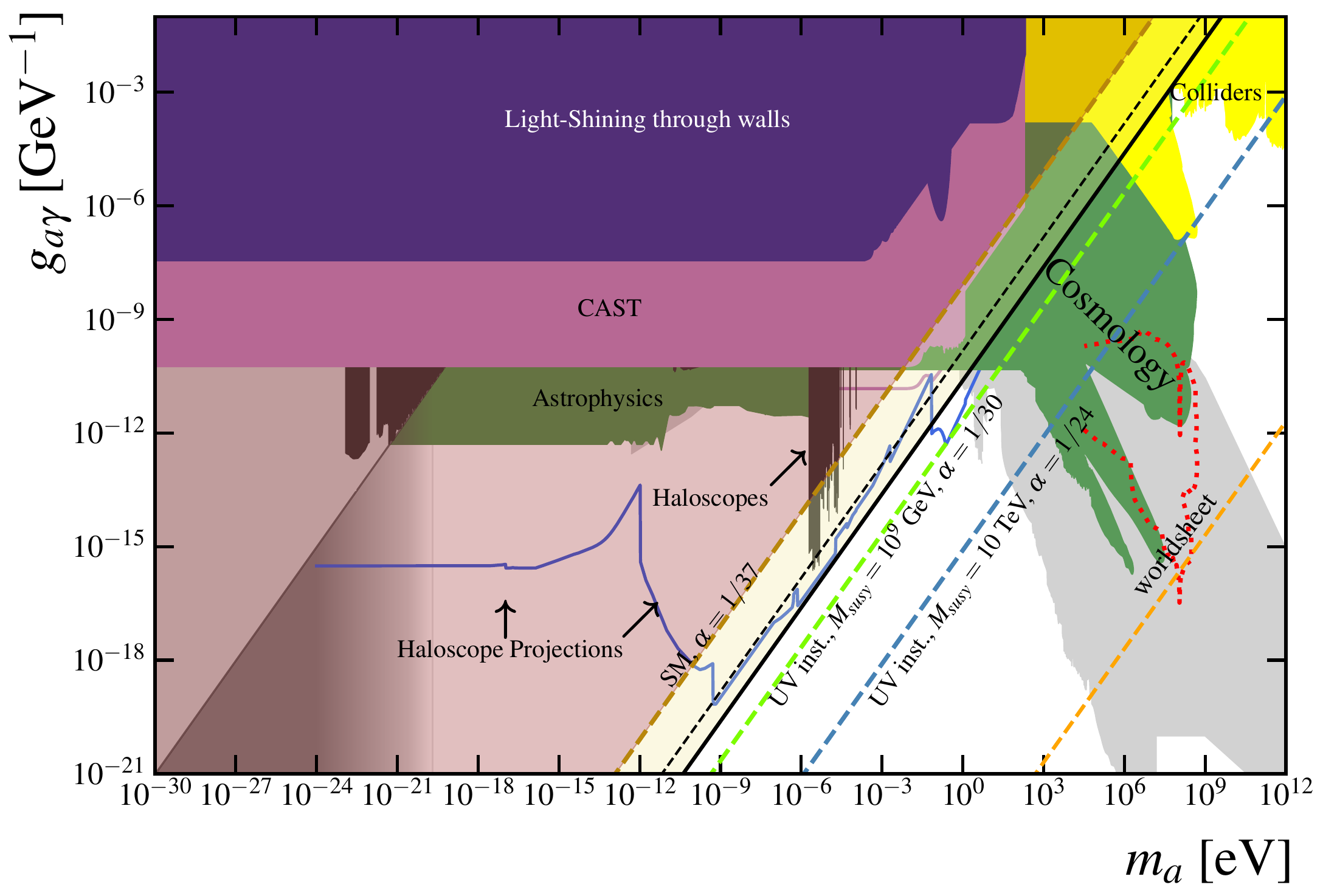}
    \caption{Same as  Figure~\ref{fig:ParameterSpace} but including negative threshold corrections, $\widehat{\Delta}_0<0$. As an example, we have taken $5\widehat{\Delta}_0 = \frac{3}{4}\frac{k_1^{(2)}}{\alpha_{\rm GUT}}$, that is 75\% of the tree-level contribution and fixed the $E_8^{(2)}$ and $U(1)_1$ instanton actions accordingly, see Eq.~\eqref{eq:inst_actions_negative_threshold}. The $U(1)_1$ instanton has the smallest action implying that the $\varphi$ has $g_{\varphi\gamma}/m_\varphi\ll g_{\theta_2\gamma}/m_{\theta_2}$ (see text for details). For this reason, we focus on the prediction of the coupling-to-mass ratio for $\theta_2$.  Situations with $\alpha_{\rm GUT}^{-1}=24$ (blue) and $\alpha_{\rm GUT}^{-1}=30$ (light green) lead to $g_{\theta_2\gamma}/m_{\theta_2}$  smaller than the QCD axion prediction (solid black line). For the benchmark $\alpha_{\rm GUT}^{-1}=37$, set by the gold shaded region, the ALP $\theta_2$ can lie above the QCD band albeit by a small margin. Smaller threshold corrections will lead to even smaller coupling-to-mass ratio for  both ALPs. 
    For simplicity, lines corresponding to heavy ALPs from confining groups are not included. The worldsheet instanton line (orange dashed) is kept for reference. See text for the case with positive threshold corrections.}
     \label{fig:ParameterSpace_with_threshold_correction}
\end{figure}

In Section \ref{sec:unif_gauge_coupling} we have pointed out, following \cite{Blumenhagen:2006ux}, that in models with non-standard hypercharge embedding the observed gauge couplings can be obtained in principle if the threshold corrections satisfy Eq.~\eqref{DeltaYalphaGUT}.
By construction, such models are at best on the verge of being under perturbative control.
 To be complete, we nonetheless analyse the potential implications of such scenarios for the coupling-to-mass ratio of ALPs. As we will see, these corrections have an important impact on the estimates of the ratio $g_{a\gamma}/m_a$ for both ALPs, $\theta_2$ and $\varphi$. 
  The reason is that the UV instanton action receives sizeable corrections in the case where $\Delta_Y$ is large, see Eqs.~\eqref{eq:YM_corrected_with_threshold} and \eqref{f11def}.

There are two different  instantons that are relevant for the discussion. On the one hand we have small $E_8^{(i)}$ instantons. For $E_8^{(1)}$, these correspond to UV instantons of the YM part of the visible sector gauge group and have the usual expression in terms of the unified gauge coupling (cf. (\ref{alphac-def})), 
\begin{equation}
    S_{E_8^{(1)}}^{\rm UV} = \frac{2\pi}{ \alpha_{\rm{YM, }\,0}}=\frac{2\pi}{\alpha_{\rm GUT}}\,.
\end{equation}
Note that the universal threshold corrections $\Delta^{\rm YM}_i$ are already included in the definition of $\alpha_{\rm GUT}$. 
Separately, there exist UV instantons that belong to $E_8^{(2)}$. These instantons have a straightforward, standard interpretation in cases where a YM subgroup survives in the 4d EFT (even if eventually Higgsed). In line bundle models where the 4d gauge symmetry is simply $\prod_m U(1)_m$ they can be described as instantons associated to the linear combination $U(1)_D$, as we discussed in Section \ref{sec:SmallInstantons}. This instanton has an action given in terms of the GUT gauge coupling and the threshold corrections,
\begin{align}
    S_{E_8^{(2)}}^{\rm UV} = \frac{2\pi}{\alpha_{\rm{YM, }\,1}}=2\pi\left ( \frac{1}{\alpha_{\rm GUT}}+\Delta_{1}^{\rm YM}-\Delta_{0}^{\rm YM}\right )=
     2\pi\left ( \frac{1}{\alpha_{\rm GUT}}+\frac{1}{k_1^{(2)}}\left (\Delta_Y- 5 \widehat\Delta_0\right ) \right )\,,
\label{eq:corrected_instanton_action_YM}\end{align}
where we have used that the hypercharge threshold was computed in (\ref{eq:threshold_hypercharge})  as
$\Delta_Y = \Delta_{Y, \rm ab} + + k_1^{(2)}(\Delta_{1}^{\rm YM}-\Delta_{0}^{\rm YM})$
and we are recalling the definition (\ref{def-hatdeltai}), and the result $\Delta_{Y, \rm ab}=5 \widehat{\Delta}_0 $. We furthermore recall from Section \ref{sec:unif_gauge_coupling} that for simplicity we assume that $U(1)_Y$ receives contributions only from two $U(1)$ factors, generalisatons being obvious.
 The instanton (\ref{eq:corrected_instanton_action_YM}) couples to $\theta_2$ and is responsible for generating part of the ALP potential, $V(\theta_2,\varphi)\supset-\kappa_{\theta_2}\Lambda_{\theta_2}^4\cos\theta_2$.

On the other hand, as argued in Section \ref{sec:SmallInstantons}, there are instantons associated to all of the ${U(1)}_{m_i}$ gauge symmetries, normalised as in (\ref{def-U(1)hat}). For simplicity here we only describe the instantons associated to ${U(1)}_0$ and ${U(1)}_1$ since these are the relevant symmetries for hypercharge and, therefore, the associated instantons break the shift symmetry of the non-universal axion $\varphi$. Their actions can be written in terms of the GUT coupling and the one-loop threshold corrections,
\begin{align}\label{eq:corrected_instanton_action_U(1)_0}
 & S_{{U(1)}_{0}}^{\rm UV} = \frac{2\pi}{ \widehat{\alpha}_{ {U(1)},\,0}}=2\pi\left ( \frac{1}{\alpha_{\rm GUT}}+\widehat{\Delta}_0\right )\,,\\&
    S_{{U(1)}_1}^{\rm UV} = \frac{2\pi}{\widehat{\alpha}_{{{U(1)}, }\,1}}=2\pi\left ( \frac{1}{\alpha_{\rm GUT}}+\Delta_{1}^{\rm YM}-\Delta_{0}^{\rm YM} + \widehat{\Delta}_1 \right )=2\pi\left ( \frac{1}{\alpha_{\rm GUT}}+\frac{1}{k_1^{(2)}}\left (\Delta_Y- \frac{10}{3}\widehat{\Delta}_{0}\right )\right )\,,
\label{eq:corrected_instanton_action_U(1)}\end{align}
where we used $\widehat{\Delta}_1 = \frac{5}{3} \widehat{\Delta}_0/k_1^{(2)}$.
With all the relevant instantons and their actions described, we now split the discussion in two different regimes according to the sign of the abelian thresholds, $\widehat{\Delta}_i$. In the case of interest, there is a single linearly independent line bundle, which implies there is only one independent threshold correction which for convenience we choose to be $\widehat{\Delta}_0$.  We will assume that $\Delta_Y<0$ so that threshold corrections improve the tree-level gauge coupling values. \\\\
\noindent\textbf{\underline{Positive threshold corrections $\widehat{\Delta}_0>0$}}\\
Let us first consider positive threshold corrections, $\widehat{\Delta}_0>0$, as realised in the class of models described in earlier studies~\cite{Blumenhagen:2006ux}, including those presented in Appendices~\ref{App:3gen}, \ref{app_ksmaller1}. In this case gauge coupling unification cannot be achieved solely by threshold corrections since for $\widehat{\Delta}_0 > 0$ it is required that $\Delta_Y>-k_1^{(2)}\alpha_{\rm GUT}^{-1}$ in order for the instanton actions to remain positive. As before, we parametrise the departure from unification as $\varepsilon=k_1^{(2)}\alpha^{-1}_{\rm GUT}+\Delta_Y$, and assume it is small, $\varepsilon \ll \alpha^{-1}_{\rm GUT}$. Altogether this yields an upper bound on the corrections, 
$\frac{10}{3}\widehat{\Delta}_0\lesssim \varepsilon$, which in turn implies that both actions are small, $S_{E_8^{(2)}}^{\rm UV}$, $S_{U(1)_1}^{\rm UV}\lesssim O(\varepsilon)$. Hence in this scenario, both ALPs, $\theta_2$ and $\varphi$, are heavy (with masses close to the KK scale) and satisfy Eq.~\eqref{eq:coupling_mass_relation}.\\

\noindent\textbf{\underline{Negative threshold corrections $\widehat{\Delta}_0<0$}}\\
In this case, we can engineer scenarios where the thresholds fix the prediction for the hypercharge gauge coupling, $\Delta_Y=-k_1^{(2)}\alpha_{\rm GUT}^{-1}$.
Assuming this, from Eqs.~\eqref{eq:corrected_instanton_action_YM}, \eqref{eq:corrected_instanton_action_U(1)} we see that the actions are proportional to the threshold corrections,  
\begin{align}\label{eq:inst_actions_negative_threshold}
    &S_{E_8^{(2)}}^{\rm UV}=-2\pi \frac{5 \widehat{\Delta}_0}{k_1^{(2)}}\,,\\&
    S_{{U(1)}_1}^{\rm UV}=-2\pi \frac{10}{3} \frac{\widehat{\Delta}_0}{k_1^{(2)}}\,.
\end{align}
On the other hand the $U(1)_0$ instanton action is necessarily smaller than the tree-level prediction, $S_{U(1)_0}^{\rm UV}=2\pi (\alpha_{\rm GUT}^{-1}+\widehat{\Delta}_0)$.

For models under perturbative control the non-universal one-loop threshold for the hypercharge gauge coupling must not exceed the tree-level contribution, $|\Delta_{Y, ab}|\lesssim k_1^{(2)}/\alpha_{\rm GUT}$, which in turn implies
\begin{equation}
    5\widehat{\Delta}_0\lesssim \frac{k_1^{(2)}}{\alpha_{\rm GUT}}\,.
\end{equation}
This is an important result which indicates that even for negative threshold corrections, in perturbative heterotic models where the IR value of gauge couplings is reproduced,
 the instanton actions are
\begin{equation}
    S_{U(1)_1}^{\rm UV}<S_{E_8^{(2)}}  \lesssim 2\pi/\alpha_{\rm GUT}\,,
\end{equation}
and both ALPs $\theta_2$ and $\varphi$ have ratios $g_{a\gamma}/m_a$ smaller than the leading order predictions in  Figure~\ref{fig:ParameterSpace}. 

The smallness of these actions, and therefore of the ratios $g_{a\gamma}/m_{a}$, depends on the specifics of the model. In  Figure~\ref{fig:ParameterSpace_with_threshold_correction} we show the effects for a theory where the threshold correction is 75\% of the tree-level contribution, $5\widehat{\Delta}_0=\frac{3}{4}\frac{k_1^{(2)}}{\alpha_{\rm GUT}}$. Any threshold correction smaller than this value will lead to coupling-to-mass ratios smaller than the ones shown. Furthermore, since the $U(1)_1$ instanton action is 2/3 of the $E_8^{(2)}$ instanton, the ALP $\varphi$ has $g_{\varphi\gamma}/m_\varphi\ll g_{\theta_2\gamma}/m_{\theta_2}$ and always satisfies Eq.~\eqref{eq:coupling_mass_relation}.

Setting aside for a moment all reservations against the resulting loss of perturbative control, we can also analyse the cases where the one-loop threshold is larger than the tree-level prediction, $|\widehat{\Delta}_i|>1/\alpha_{\rm GUT}$. This hypothetical scenario can lead to large actions $S_{E_8^{(2)}}^{\rm UV}$, $S_{U(1)_1}^{\rm UV}\gtrsim 2\pi/\alpha_{\rm GUT}$, and the ALP $\theta_2$ can remain light. The non-universal ALP $\varphi$ will obtain its mass from the $U(1)_0$ instanton, with an action smaller than the tree-level prediction, 
 $S_{U(1)_0}^{\rm UV}< 2\pi/\alpha_{\rm GUT}$,
 and is expected to gain a large mass. 

However, as discussed in Section \ref{sec_theorieswithlargethresholds}, this has important implications. In particular,  Eq.~\eqref{eq:corrected_instanton_action_U(1)_0} suggests that a negative $\widehat{\Delta}_0$ with $|\widehat{\Delta}_0|\geq |\Delta_Y|$ leads to a cancellation in the UV instanton action from the first (visible sector) $E_8$, $S_{U(1)_0}^{\rm UV}\sim O(1)$, that alters the potential of the QCD axion, $\theta_1$. Such large instanton contributions will in general spoil the QCD axion solution to the strong CP problem.

To summarise, heterotic models with a light ALP above the QCD line can only exist in scenarios in which perturbative control is lost. The requirements for such models include negative abelian threshold corrections (see Appendix~\ref{App:3gen} for a discussion).
 The corrections should be large enough so that 
$\Delta_Y-\frac{10}{3}\widehat{\Delta}_{0}\geq 0$ together with the constraint $\Delta_Y\approx -\frac{k_1^{(2)}}{\alpha_{\rm GUT}}$. Such models therefore leave the regime of theoretical control and further non-perturbative effects are expected to become important to reliably determine axion properties such as the ratio $g_{a\gamma}/m_a$.

\section{Discussion and phenomenological implications}\label{sec:conclusions}
In this article we have studied the coupling-to-mass ratio, $g_{a\gamma}/ m_a$, for axions in heterotic string compactifications with non-standard embedding of the SM \cite{Blumenhagen:2006ux}. In such constructions it is possible to embed the SM hypercharge, $U(1)_Y$, non-trivially into both $E_8$ factors, break the second $E_8$ down to $\prod_m U(1)_m$, and obtain a theory without chiral exotics. 
 Since $U(1)_Y$ consists of a combination of $U(1)$s in both $E_8$ groups, there arise two ALPs, $\theta_2$ and $\varphi$, that couple to photons without coupling to gluons. This construction offers, in principle, the ingredients to violate the bound in Eq.~\eqref{eq:coupling_mass_relation}, found for perturbative heterotic models where the SM is embedded into a single $E_8$. The main result of this work is that even though slightly modified, a similar bound remains in place  within the regime of perturbatively controlled heterotic string theory.

\subsection{Main results}

Due to the coupling to gauge bosons, a shift-symmetry breaking potential $V(\theta_2, \varphi)$ is generated for the ALPs. This depends on the concrete gauge background which determines the unbroken gauge group from the second $E_8$ that partially contains $U(1)_Y$. If a non-abelian confining group $G_{\rm hidden}$ survives in the 4d EFT, one of the ALPs, $\theta_2$, is always much heavier than the QCD axion and satisfies \eqref{eq:coupling_mass_relation}. For the remaining ALP, $\varphi$, and generally in cases where only $U(1)$s survive after the embedding of the gauge background, the relevant non-perturbative objects generating axion masses are worldsheet instantons and Euclidean NS5-branes. The former break the shift symmetries of the model dependent axions $b_i$ and $\tilde b_r$ coming from $B_2$ and from the NS-brane sector, respectively, but leave the model independent axion $a$ unaffected. In some particular cases where 2-cycles are small, for example when $g_s^2\ll 1$ (see Section \ref{sec:small_gs}) or if $h_{1,1}\gg O(1)$, the typical worldsheet instanton action is $S_{\rm ws}\sim O(1)$ and one expects the model dependent axions to be heavy with $g_{b_i \gamma}/m_{b_i}$ much smaller than the QCD axion prediction. On the other hand, NS5-brane instantons induce shift symmetry breaking effects that can be understood in terms of small gauge instantons at the string scale. Since they are suppressed by the inverse gauge couplings, these effects hence offer more model independent constraints. 
 Systematically analysing their impact on  the ALP coupling-to-mass ratio is the key of our analysis.

In the theories with non-standard hypercharge embedding that we consider (based on the models studied in \cite{Blumenhagen:2006ux}) the ratio $g_{\theta_2\gamma}/m_{\theta_2}$ and  $g_{\varphi\gamma}/m_{\varphi}$ can be larger than the QCD axion prediction but, crucially, an upper bound can be estimated once the supersymmetry breaking scale and $\alpha_{\rm GUT}$ are fixed (assuming no additional matter beyond the three chiral families of the SM; cases with additional matter were studied in Section \ref{sec:intermediate_states_ALP_potential}). 
The \textit{tree-level} or leading order predictions for the coupling-to-mass ratio for the ALP and the QCD axion for different values of $m_{\rm susy}$ and $\alpha_{\rm GUT}$ have been studied in Section \ref{sec:ALP_potential} and the results are shown in  Figure~\ref{fig:ParameterSpace}. However, as discussed in section~\ref{sec:unif_gauge_coupling}, these predictions correspond to models which are not phenomenologically viable due to the incorrect prediction of the gauge couplings and $\sin^2\theta_w$ at the GUT scale.

To refine our estimates, in Section \ref{sec_thresholds}, we have studied how the predictions for $g_{\theta_2\gamma}/m_{\theta_2}$ or $g_{\varphi\gamma}/m_{\varphi}$ are modified once the different mechanisms available to obtain the correct gauge couplings are taken into consideration. These can be separated in two large classes. In the first class of models, the threshold corrections are small, $\Delta_Y$,$\widehat{\Delta}_i\ll 1$.
 In this case, the correct gauge couplings can be obtained in presence of intermediate scale matter with $SU(2)_w$ and/or $SU(3)_C$ charges, or because the level of embedding of hypercharge $U(1)_Y$ into the second $E_8$ is engineered to be small, $k_1^{(2)}\ll 1$. In the first case, the new charged states end up decreasing the relevant UV instanton action and the coupling-to-mass ratios for the ALPs are drastically reduced with respect to the leading order estimates. In these theories we conclude there is no light ALP above the QCD axion line. For $k_1^{(2)}\ll 1$ we find that $g_{\theta_2\gamma}/m_{\theta_2}$ and $g_{\varphi\gamma}/m_{\varphi}$ are also reduced due to the ALP coupling to photons being proportional to $(k_1^{(2)})^2$, see Eq.~\eqref{eq:g_a_in_small_k1_2}. An example with $k_1^{(2)}=1/9$ is shown in  Figure~\ref{fig:ParameterSpace_with_small_k1}. We remark that in either of these scenarios with small threshold corrections, we also expect sizeable effects from worldsheet instantons. For this reason our results are conservative in this limit. Including worldsheet instanton effects systematically and studying how their action $S_{\rm ws}$ scales with $h_{11}$ is an interesting direction left for future work. 

In the second class of models, sizeable threshold corrections  from heavy string modes improve the unification of gauge couplings if $\Delta_Y\approx -k_1^{(2)}/\alpha_{\rm GUT}$, see Eq.~\eqref{DeltaYalphaGUT}. In this case the ALP coupling-to-mass ratio depends on the sign of the abelian threshold corrections given by Eq. (\ref{Detltami-gen}). For $\widehat{\Delta}_i\geq 0$, which includes the large class of models in \cite{Blumenhagen:2006ux}, there is no light ALP because large cancellations in the relevant UV instanton actions result in $S_{U(1)_1}^{\rm UV}\sim S_{E_8^{(2)}}^{\rm UV}\sim O(1)$. Only if the abelian threshold corrections are negative and large, $\frac{10}{3}|\widehat{\Delta}_0|\geq |\Delta_Y|$, can a light ALP appear. 
 However, as discussed in Section \ref{sec_theorieswithlargethresholds}, this has important implications as it leads to UV instantons with small action that may spoil the QCD axion solution to the strong CP problem. More importantly, in situations with such large threshold corrections, perturbation theory in $g_s$ is lost and a more reliable description is offered by M- or F-theory. 
 In this sense, the statement upholds that a sizeable violation of the bound \eqref{eq:coupling_mass_relation} is not possible within the \textit{perturbative} heterotic string.

A prototypical situation where the threshold corrections are large, but still (slightly) subdominant with respect to the tree-level contribution is shown in Figure  \ref{fig:ParameterSpace_with_threshold_correction}. As this example illustrates, in models where the one-loop string threshold corrections are smaller than $\sim 75\%$ of the tree-level contribution, the ALPs $\theta_2 , \varphi$ have a $g_{a\gamma}/m_{a}$ smaller than the QCD axion prediction, even if $m_{\rm susy}$ is close to the KK scale and $\alpha_{\rm GUT}\sim 1/37$.

Summarizing, we have obtained the important result that the \textit{tree-level} estimates in  Figure~\ref{fig:ParameterSpace} are conservative upper bounds to the ratio $g_{\theta_2\gamma}/m_{\theta_2}$ and $g_{\varphi\gamma}/m_{\varphi}$ for each of the values of $m_{\rm susy}$ and $\alpha_{\rm GUT}$. Any of the mechanisms addressing the issue of the gauge coupling unification in perturbatively controlled models with non-standard $U(1)_Y$ embedding decrease the leading order predictions for the coupling-to-mass ratio $g_{a\gamma}/m_a$. 

\subsection{Implications for phenomenology and cosmology}

The described bounds to $g_{a\gamma}/m_a$ have important phenomenological implications. There exist several experimental probes which are sensitive to light axions above the QCD line including laboratory, astrophysical and cosmological probes (see  Figure~\ref{fig:ParameterSpace}). A prototypical example of an axion with $g_{a\gamma}/m_a$ larger than the QCD axion is that underlying cosmic birefringence~\cite{Carroll:1989vb,Harari:1992ea,Carroll:1998zi}. This effect describes the rotation of the polarization angle of the cosmic microwave background (CMB) as it travels through space in presence of an ultralight axion; this axion must be coupled to photons and have a mass between today's and the CMB epoch Hubble scale, $H_0\lesssim m_a\lesssim H_{\rm CMB}$. 
 The bounds we have obtained in any of the phenomenologically viable theories, however, are incompatible with such a signal. Therefore, if this or similar signals requiring an axion coupled to photons with $g_{a\gamma}/m_a\gg 0.01 \text{ GeV}^{-2}$ are found in a near future, they would have tremendous implications for string model building. In particular, discovering such an axion coupled to photons would tell us that perturbative heterotic string theory and type-I string theory (with the SM embedded into $SO(32)$) cannot be the UV completion of the SM. This is particularly important given the cosmic birefringence hints in~\cite{Minami:2020odp,Eskilt:2022cff,Diego-Palazuelos:2023mpy,Zhang:2024dmi,Diego-Palazuelos:2025dmh}.
  On the other hand, finding an ALP between the QCD line and the conservative upper bound shown in Figure \ref{fig:ParameterSpace} might still be compatible with perturbative heterotic models, but only for certain ranges of the supersymmetry breaking scale. This illustrates the expectation that to ultimately constrain the UV completion of the SM, a combination of several observations is required. We find the predictive power of light ALP searches for fundamental physics, as exemplified by our study, quite intriguing.

Conversely, our analysis suggests that a quite generic implication of perturbative 
 heterotic models both for standard and non-standard embeddings is the presence of heavy ALPs coupled to photons.\footnote{We thank Ben Safdi for insisting that this is not necessarily a nightmare scenario.} In field theory constructions the existence of these particles might appear arbitrary. By contrast, in the heterotic axiverse they appear naturally. 
  Interestingly, unlike type-II string models, where the coupling of most axions in the axiverse is suppressed by localisation of the gauge sectors in the gravitational bulk \cite{Gendler:2023kjt}, we expect that a sizeable part, if not most, of axions in the heterotic axiverse have \textit{unsuppressed} couplings to gauge bosons (scaling as $\sim O(1)/f_a$) via the Green-Schwarz mechanism~\eqref{eq:GS}. These heavy ALPs may be abundantly produced in the early universe and leave an interesting imprint in the early universe cosmology. It would be important to perform more quantitative studies of this scenario.

\section*{Acknowledgements}
We thank Josh Benabou, Ben Safdi, Nils Tomforde, Cumrun Vafa for useful discussions. MR thanks Prateek Agrawal and Michael Nee for amazing collaborations in related projects.
We thank the organizers of the workshop \textit{Dark World to Swampland 2024}, as well as IBS in Daejeon, for their hospitality, where this project was initiated. T.W. thanks the Simons Foundation for hospitality at the Simons Summer Workshop 2025.
The work of  T.W. is supported in part by Deutsche Forschungsgemeinschaft under Germany’s Excellence Strategy EXC 2121 Quantum Universe 390833306, by Deutsche Forschungsgemeinschaft through a German-Israeli Project Cooperation (DIP) grant “Holography and the Swampland” and by Deutsche Forschungsgemeinschaft through the Collaborative Research Center 1624 “Higher Structures, Moduli Spaces and Integrability.” This article is based in part upon work from COST Action COSMIC WISPers CA21106, supported by COST (European Cooperation in Science and Technology).

\appendix 

\section{A 3-generation model with non-standard hypercharge embedding} \label{App:3gen}
In this appendix we present a model with three chiral generations of SM matter and no chiral exotics, where hypercharge is embedded into both $E_8$ factors and where the second $E_8$ is broken to abelian gauge factors. This model is based on the construction \cite{Blumenhagen:2006ux}, to which we refer for more details. 

 As noted in Section \ref{sec:constrain_line_bundle}, for phenomenological reasons, the chiral index 
 \begin{equation} \label{chiqi-1}
\chi(1_{\vec{q}}) =
\frac{1}{6} \int_X\left(\sum_{m_2 = 1}^8 q^{m_2}_i c_1(L_{m_2})\right)^3 + \sum_{m_2 = 1}^8   \frac{1}{12} q^{m_2}_i \int_X  c_1(L_{m_2}) \wedge c_2(X)\,,   
\end{equation}
of the singlets 
 charged under the abelian gauge sector from the second $E_8$ must vanish for all those states which are charged under the specific linear combination $U(1)_Y$ that corresponds to the SM hypercharge.
 
 Let us denote by $U(1)_{m_2}$ for $m_2 \in {\cal I}$ those $U(1)$s which appear in this linear combination.
 Then the condition of vanishing chiral index for singlets with hypercharge can be easily ensured when the Calabi-Yau threefold $X$ admits an elliptic fibration over a complex surface $B_2$ and if the first Chern classes of the line bundles $L_{m_2}$ for $m_2 \in {\cal I}$ are pulled back from $B_2$ as
\begin{equation} \label{pullbackLa-1}
c_1(L_{m_2}) = \pi^\ast(D_{m_2})   \,, \qquad D_{m_2} \in H^{1,1}(B_2) \,, \qquad m_2 \in {\cal I} \,.
\end{equation}
Here $\pi$ is the projection of the elliptic fibration and $D_{m_2}$ is a divisor class on the base $B_2$.
Since $c_1(L_{m_2})$ is the pullback of a class from the base $B_2$, these bundles $L_{m_2}$ automatically satisfy the relations 
\begin{equation}
\int_X c_1(L_{m_2}) \wedge c_1(L_{n_2})  \wedge c_1(L_{p_2})  = 0\,, \quad {\rm \forall} \,  {m_2},{n_2}, {p_2} \in {\cal I}  \,.
\end{equation}
To evaluate the condition on vanishing chirality, we further note that for an elliptic fibration with section $\sigma$, the second Chern class $c_2(X)$ is given by
\begin{equation} \label{c2X-gen}
c_2(X) = 12 \sigma \wedge c_1(B_2) + n [F] \,,
\end{equation}
where $[F]$ denotes the class of the fiber and for smooth Weierstrass models, $n= \int_{B_2} 11 c_1^2(B_2) + c_2(B)$. Since $\int_X [F] \wedge \pi^\ast(D_a) =0$, this implies $\int_X c_1(L_{m_2})  \wedge c_2(X) = 12 \int_{B_2} c_1(B_2) \wedge  D_{m_2}$. Hence $\chi(1_{\vec{q}}) =0$ is achieved, in particular, if  
\begin{equation} \label{c1Bc1Lcondition}
\int_{B_2} c_1(B_2) \wedge  D_{m_2} = 0 \, \quad \forall \, {m_2} \in {\cal I}\,.
\end{equation}
 
 Specifically, let us pick as basis $B_2$
 the del Pezzo surface $dP_4$, which is a projective surface $\mathbb P^2$ blown up in four points. 
 Its second cohomology group is spanned by the hyperplane class $l$ inherited from  $\mathbb P^2$ as well as four blow-up curve classes
 $E_1$, $E_2$, $E_3$, $E_4$ 
 with intersection form 
 \begin{equation} \label{intform1}
 l\cdot l=1 \,, \qquad  E_i \cdot E_j = - \delta_{i,j}\,, \qquad  l\cdot E_i = 0 \,.
\end{equation}
 The first Chern class of $dP_4$ is given by 
 \begin{equation}\label{c1dP4}
 c_1(B_2) = 3 l - E_1 - E_2 -E_3 - E_4 \,, \qquad {\rm with} \quad \int_{B_2} c_1^2(B_2) =: c_1(B_2) \cdot c_1(B_2) = 5 \,.
\end{equation}
Together with $c_2(dP_4) = 7$, this gives for the second Chern class (\ref{c2X-gen}) of the elliptic fibration $X$, which we construct as a smooth Weierstrass model over $B_2$, 
\begin{equation} \label{c2X-app}
 c_2(X) = ( 36 l - 12 E_1 - 12 E_2 - 12 E_3 - 12 E_4) \sigma + 62 [F] \,,
\end{equation}
with $F$ denoting the class of the generic elliptic fiber.\footnote{Here and in the sequel we sometimes omit the wedge product symbol to make the notation lighter. Furthermore we omit the pullback symbol and denote e.g. by the expression in the bracket in (\ref{c2X-app}) also the element in $H^2(X)$ obtained by pulling back the class that is written from $B_2$.}

From the discussion around (\ref{SMemb-1}) we recall that we aim to embed the direct sum \begin{equation}
 W_1 = V \oplus L_0^{-1}
\end{equation}
 of a $U(5)$ vector bundle $V$ and a line bundle $L_1^{-1}$ with  
 \begin{equation}
 c_1(V) = c_1(L_0)
 \end{equation}
 into the first $E_8$ in such a way that the commutant of its structure group is $SU(3) \times SU(2) \times U(1)_0$.
  The precise embedding is chosen by identifying the generator of the structure group of the line bundle $L_0^{-1}$ with the $U(1)$ generator $(1,1,1,1,1,-5)$ of the total structure group $SU(5) \times U(1)$ of $W_1$.
In the second $E_8$, we embed the direct sum of line bundles 
\begin{equation}
W_2= \oplus_{m_2=1}^8 L_{m_2} 
\end{equation}
with $c_1(W_2)=0$ to realise a breaking pattern of the form (\ref{hiddenpattern}).
 The $({\bf 248}, 1) + (1, {\bf 248})$ of $E_8 \times E_8$ decomposes into a sum of representations under the structure group ${ SU(5)}$ of $V$ as well as the visible ${SU(3)} \times {SU(2)} \times { U(1)}_0 \times \prod_{m_2} {U(1)}_{m_2}$. The part of this spectrum with charge under one of the abelian gauge group factors is listed in the left column of Table \ref{tab:SM-spectrum}. 
 The chiral index of the various vector bundles $U$ that appear in this table is given by
 \begin{equation}
\chi(U) = \int_X {\rm ch}_3(U) + \frac{1}{12} \int_X c_1(U) \wedge c_2(X) \,.
 \end{equation}

 In addition, the massless spectrum comprises the gauge multiplets for the gauge group factors and a number of massless singlet moduli.

Concretely, for $V$ we can take a vector bundle obtained via the spectral cover construction on $X$.
 The Chern classes of such a bundle is uniquely determined by specifying two classes, $c_1(\zeta)$ and $\eta$, in $H^2(B_2,\mathbb Z)$, a number $\lambda \in \mathbb Q$ and an integer $q$. Setting $q=0$ for simplicity, the Chern characters of such a bundle take the form 
 \begin{eqnarray}
{\rm ch}_1(V) &=& 5 c_1(\zeta) \,, \\
{\rm ch}_2(V) &=& - \eta  \sigma + \frac{5}{2} c_1^2(\zeta) - ( - 5 c_1(B_2)^2 + \frac{5}{2}(\lambda^2 - \frac{1}{4}) \eta (\eta - 5 c_1(B_2))) \,, \\
{\rm ch}_3(V) &=& \lambda \eta (\eta - 5 c_1(B_2)) - \eta c_1(\zeta) \,.
 \end{eqnarray}
For more details of the spectral cover construction in the present context we refer to \cite{Blumenhagen:2006ux}.
 Note in particular that the choice $q=0$ implies that
 $c_1(V)$, and hence also $c_1(L_1)$, is a pullback from the base of the elliptic fibration.
For example, the choice
\begin{equation}
\eta = 15 l - 5 E_1 - 5 E_2 - 3 E_3 - 5 E_4 \,, \qquad \lambda = \frac{1}{2} \,, \qquad c_1(\zeta) = - E_1 + E_4
\end{equation}
 
leads to a stable $U(5)$ bundle with
\begin{eqnarray}
c_1(V) &=& 5 c_1(\zeta) = 5  (-E_1 + E_4)  \,, \\
c_2(V) &=& -5 c_1(B_2)^2 + \eta \sigma + 10 c_1(\zeta^2) =    \sigma (15 l - 5 E_1 - 5 E_2 - 3 E_3 - 5 E_4) - 45 [F] \,.
\end{eqnarray}
This bundle is one of the examples constructed in \cite{Blumenhagen:2006ux} (see Table 6 therein)
 which give rise to a 
 visible sector with three chiral families of SM matter. The latter statement follows by evaluating the chiral index of the matter states listed in Table \ref{tab:SM-spectrum}.

\begin{table}[t!]
 \centering
    \begin{tabular}{|c||c|}
    \hline  Representation &  $ {\rm Chiral \, index}$  \\ \hline \hline
       $({\bf 5} \, | \,  {\bf 3}, {\bf 2},1)_{1, \vec{0}}$  & $\chi(V) = 3$ \\ \hline 
$({\bf 1} \, | \,  {\bf 3}, {\bf 2},1)_{-5, \vec{0}}$ & $\chi(L_0^{-1}) = 0$  \\ \hline 
$({\bf 10} \, | \,  {\bf \bar 3}, {\bf 1},1)_{2, \vec{0}}$  & $\chi(\bigwedge^2 V) = 3$ \\ \hline
$({\bf 5} \, | \, {\bf \bar 3}, {\bf 1},1)_{-4, \vec{0}}$  & $\chi(V \otimes L_0^{-1}) = 3$ \\ \hline   
$({\bf 10} \, | \, {\bf 1}, {\bf 2},1)_{-3, \vec{0}}$  & $\chi(\bigwedge^2 V \otimes L_0^{-1}) = 3$ \\ \hline
$({\bf 5} \, | \, {\bf 1}, {\bf 1},1)_{6, \vec{0}}$  & $\chi(V \otimes L_0) = 3$ \\ \hline \hline
$({\bf 1} \, | \, {\bf 1}, {\bf 1},1)_{0, \vec{q}_i}$  & $\chi(\prod_{m_2} L_{m_2}^{q^{m_2}_i}) = 0$ \\ \hline
 \end{tabular}
    \caption{Spectrum of SM construction with hypercharge embedded into both $E_8$ factors. The left corner gives the representation under  $SU(5)\,  |  \, \,  {SU(3)} \times  { SU(2)} \times  { U(1)}_0  \times \prod_{m=1}^8 U(1)_m$, where $SU(5)$ refers 
     to the non-abelian part of the structure group of the bundle $V$ and is not visible as a representation in the effective field theory. All but the last line contain states from the decomposition of the first $E_8$. }
    \label{tab:SM-spectrum}
\end{table}

 To engineer the second $E_8$ bundle, we can therefore pick, for example,
\begin{equation}
c_1(L_{1}) =   5(-E_1 + E_4) \,, \quad c_1(L_{m_2}) =   E_1 - E_4\,, \quad    m_2 = 2, \ldots 7 \,, \qquad c_1(L_8) =  -E_1 + E_4\,,
\end{equation}
satisfying the constraint $c_1(W_2) =0$.
These line bundles all obey the condition (\ref{c1Bc1Lcondition}) that guarantees absence of chiral exotics.

The Bianchi identity for the gauge background from both $E_8$ factors takes the form
\begin{equation}
(-c_2(V) + c_1^2(L_0)) + \frac{1}{2} \sum_{m_2=1}^8 c_1^2(L_{m_2}) - [W] = - c_2(X) \,,
\end{equation}
where $[W]$ denotes the class of a curve on $X$ wrapped by a heterotic 5-brane.
For the above model, the Bianchi identity is satisfied for
 $[W] = 25 F + (21 l - 7 E_1 - 7 E_2 - 9 E_3 - 7 E_4) \sigma$.
 This class is effective, as required for a model that is supersymmetric and stable at the string scale.
 
 Finally, the D-term condition on the gauge bundles reduces to the single constraint
 \begin{equation}
\int_X J \wedge J \wedge c_1(V) = 0 \,.
 \end{equation}
 This is because for the above choice of bundles, the $g_s$-dependent dependent correction to the D-term \cite{Blumenhagen:2005ga, Blumenhagen:2006ux} vanishes automatically.
  The D-term condition can be satisfied within the K\"ahler cone of the Calabi-Yau, $X$.

The mass matrix for the St\"uckelberg mechanism has rank one. This means that only one linear combination of the nine $U(1)$ gauge factors ${U(1)}_0$ (from the first $E_8$) and ${ U(1)}_{m_2}$, $m_2= 1, \ldots 8$, from the second $E_8$ is massive and only survives as a global symmetry, while the orthogonal ones remain unbroken. 
 For example, if we start with the decomposition (\ref{eq:248_splitting}) and continue to break $E_7$ by embedding further line bundles, the group theory factors for $U(1)_1$ (and $U(1)_0$ from the first $E_8$) are unchanged compared to (\ref{kappaeta-example1}). 
Then
\begin{equation}
U(1)_0 + 3 U(1)_1
\end{equation}
is massless and can be identified with hypercharge up to normalisation. 

Taking into account the correct normalisation as in (\ref{k12-1}), this model has $k_1^{(2)} = 1$ so that gauge coupling unification without intermediate charged matter would require large negative threshold corrections (\ref{DeltaYalphaGUT}).
Despite these phenomenological drawbacks, these $\mathcal N=1$ supersymmetric 3-generation models do serve our purpose of illustrating the possibility to break the hidden $E_8$ to an abelian gauge group without the generation of chiral states with SM charges. 

 Finally, let us note that these models exemplify an important fact. Depending on the geometry, it can indeed occur that the non-universal contributions to the abelian threshold corrections, given by (\ref{Detltami-gen}), are positive,
 \begin{equation} \label{Deltami-sign}
\Delta_{m_i} > 0 \,.
 \end{equation}
As discussed in Section \ref{sec_theorieswithlargethresholds}, this has important properties for the mass of the ALP. 
 
In the present geometric setting, (\ref{Deltami-sign}) is a direct consequence of the fact that the compactification space is an elliptic fibration over base $B_2 = dP_r$ with $r < 9$, together with the condition that $c_1(L_{m_i})$ is the pullback of a divisor $D_{m_i}$ from $B_2$, see (\ref{pullbackLa-1}), as well as  (\ref{c1Bc1Lcondition}). Indeed, on a base $B_2$ with $\int_{B_2} c_1^2(B_2) > 0$, as is the case for the del Pezzo surfaces $dP_r$ with $ r < 9$, a divisor $D$ with  $\int_{B_2} D \wedge c_1(B_2) =0$  obeys $\int_{B_2}D^2 < 0$ because the intersection pairing on the K\"ahler base $B_2$ has  signature $(1,h^{1,1}(B_2) -1)$.  Since for line bundles pulled back from $B_2$,  $c_1^2(L_{m_i}) = n [F]$ with $F$ the fiber of $X$ and $n = \int_{B_2} D_{m_i}^2$, this in turn implies that the non-universal part of the $U(1)$ thresholds $\Delta_{m_i} > 0$ inside the K\"ahler cone, see (\ref{Detltami-gen}).

This argument also shows that the sign of $\Delta_{m_i}$ is model dependent -- on elliptic fibrations over base spaces with $\int_{B_2}c_1^2(B_2) < 0$,  a priori no reason forbids $\Delta_{m_i} < 0$.

\section{A flipped SU(5) model with $k_1^{(2)} < 1$} \label{app_ksmaller1}
In this appendix we demonstrate the possibility of  $k_1^{(2)} < 1$ for the level appearing in the normalization of hypercharge, (\ref{eq:hypercharge_coupling_with_threshold}).
The example is based on a flipped $SU(5) \times U(1)_X$ model.
In the visible sector we embed a $U(4) \times U(1)$ bundle $W_1 = V \oplus L_0^{-1}$ to break $E_8^{(1)} \to SU(5) \times U(1)_0$, as detailed in \cite{Blumenhagen:2006ux}, to which we refer for details. 
 We borrow the bundle $V$ constructed in \cite{Blumenhagen:2006wj} as a stable extension bundle on an elliptic fibration over $B_2 = dP_4$ with Chern classes
\begin{eqnarray}
c_1(V) &=& - E_1 + E_4 \,, \\ 
c_2(V) &=& - 13 F + \sigma (22l - 9 E_1 - 6 E_2 - 6 E_3 - 9 E_4) \,, \\
c_3(V) &=& 6 \,.
\end{eqnarray}
Note that $c_1(W_1) = 0$ requires that 
\begin{equation}
c_1(L_0) = - E_1 + E_4
  \,.
\end{equation}
This gives rise to a 3-generation flipped $SU(5)$ model.
In the second $E_8$ we can embed eight line bundles with
\begin{equation}
c_1(L_{m_2}) = \ell_1 (-E_1 + E_4) \,, \quad 1 \leq m_2 \leq 4  \,, \qquad  c_1(L_{n_2}) = \ell_1 (E_1 - E_4)  \,,\quad 5 \leq n_2 \leq 8 \,, \qquad  
\end{equation}
for some $\ell_1 \in  \mathbb Z$. 

Choosing the same class of embeddings for the second $E_8$
 as in Appendix \ref{App:3gen}, and taking into account the embedding that gives rise to flipped $SU(5)$ in the first $E_8$, one finds altogether 
 \begin{equation} \label{flippedSU5kappas}
\kappa_{0,0} = -10 \,, \quad \eta_{0,0} = 40  \,, \qquad \kappa_{1,1} = -4\,, \quad \eta_{1,1} = 4  \,.
 \end{equation}
The linear combination 
\begin{equation}
-\frac{1}{2}(U(1)_0 - \frac{5}{2 \ell_1} U(1)_1)
\end{equation}
remains massless
and plays the role of $U(1)_X$
in flipped $SU(5)$ from which hypercharge descends after GUT symmetry breaking. 
Here we used (\ref{masslessU(1)-gen}) along with the concrete group theory factors (\ref{flippedSU5kappas}) for the specific embedding, and we furthermore note that  
 $\ell_0=1$ because $c_1(L_0) = -E_1 + E_4$.
 The Bianchi identity takes the form 
\begin{equation}
- c_2(V) + c_1(L)^2 + \frac{1}{2} \sum_{m_2} c_1(L_{m_2})^2 + c_2(X) = [W] 
\end{equation}
For the above types of bundles, this gives
\begin{equation}
 [W] = (73 - 8 \ell^2) F + \sigma (14 l - 3 E_1 - 6 E_2 - 6 E_3 - 3 E_4) \,.
\end{equation}
The largest value of $\ell$ for which the coefficient in front of the fiber class $F$ is non-negative, as required for an effective 5-brane curve, is $\ell_1 = 3$.

\section{Axion couplings in the presence of NS5-branes}\label{App:non_perturbative_axions}
The presence of spacetime-filling NS5-branes modifies anomaly cancellation mechanism. This can be seen from the Bianchi identity for the 3-form $H$, which now contains new terms due to the NS5-branes,
\begin{equation} \label{Biachni-Id-gen_Appendix}
    {\rm tr}_{1} {\bar F}^2 + {\rm tr}_{2} {\bar F}^2 - {\rm tr} {\bar R}^2 = \sum_r N_r [\Gamma_r] \,,
\end{equation}
where $[\Gamma_r]$ is an effective curve class on the compactification space $X$ that is wrapped by a stack of $N_r$ spacetime-filling heterotic 5-branes, and $\bar{F}$ is the curvature 2-form of the vector bundle. As explained in the main text, the 5-branes do not change the crucial observation that, apart from the non-universal axion $\varphi$, there exist only two different linear combinations coupled to gauge bosons in the 4d EFT via the anomaly, 
\begin{equation}
    \mathcal{L} =\frac{\theta_1}{8\pi}  \tr_1F^2+\frac{\theta_2}{8\pi}\tr_2F^2 \,.
\end{equation}
The goal of this appendix is to obtain the most general expression for $\theta_{1,2}$ which is also valid in the presence of background 5-branes.

To find the modification to the axion couplings we again examine the anomaly cancellation mechanism. Due to the 5-branes, cancelling anomalies requires new counter-terms~\cite{Blumenhagen:2006ux}
\begin{align}\label{eq:new_MD_axion_coupling}    
   & S_{GS}^{\text{new, 1}} = \frac{1}{768\pi^3} \sum_r N_r \int_{\Gamma_r}B_2 \wedge \left ( \tr _1 {\cal F}^2 +\tr_2 {\cal F}^2 - \tr {\cal R}^2\right ) \,, \\&
    S_{GS}^{\text{new, 2}}= \frac{1}{64\pi^3}\sum_r N_r\int_{\Gamma_r}\tilde{B}_2^{(r)}\wedge \left ( \tr_1 {\cal F}^2 -\tr_2 {\cal F}^2 \right ) \,.\label{eq:non-pert_axion_coupling}
\end{align}
 $B_2$ is the standard 10d two-form field present in the perturbative heterotic string that couples to the fundamental string and the $\tilde B_2^{(r)}$ are additional 2-form fields living in the world-volume of space-filling $5-$branes. These new two-form fields couple to solitonic strings. In close analogy to the model dependent axions from $B_2$, when we integrate the new 2-form fields $\tilde B^{(r)}_2$ over the 2-cycles $\Gamma_r$ we obtain a new kind of axion in the 4d EFT, $\tilde b_r = \int_{\Gamma_r}\tilde B_2^{(r)}$.

From Eq.~\eqref{eq:non-pert_axion_coupling} we can see that $\tilde b_r$ couple with opposite sign to gauge sectors from the first and the second $E_8$. Equivalently, this can also be seen from the Bianchi identity for the field strength of the new two-form fields, which reads 
\begin{equation}
    d\tilde{H}^{(r)}= N_r\left ( \tr_1 {\cal F}^2 -\tr_2 {\cal F}^2\right )  \,.
\end{equation}
Since the 2-form field is self-dual in 6d, $\tilde H^{(r)} = \star\tilde H^{(r)}$, the modified Bianchi identity corresponds to the equation of motion for the axions coming from integrating $\tilde B_2^{(r)}$ over 2-cycles, $\tilde{b}_r$. In this way, one can see that their 4d couplings to gauge bosons are qualitatively equivalent to model dependent axions in the absence of NS5-branes (see discussion around Eq.~\ref{eq:different_kinds_of_axions}), which also exhibit a relative sign in their coupling to the gauge sectors coming from both $E_8$ factors. 

The contributions in Eqs.~\eqref{eq:new_MD_axion_coupling}, \eqref{eq:non-pert_axion_coupling} have to be added to the previously derived couplings of the model independent and model dependent axions to 4d gauge bosons. Altogether, the two linear combinations are 
\begin{align}
    &\theta_1 = a+\sum_in^{(1)}_ib_i+\sum_jm_j^{(1)}c_j+\sum_rN_r\tilde b_r   \,,\\&
    \theta_2 = a+\sum_in^{(2)}_ib_i+\sum_jm_j^{(2)}c_j-\sum_rN_r\tilde b_r\,.   
\end{align}
The latter terms involving the integers $N_r$ correspond to the axion couplings induced by the presence of NS5-branes, not present in the perturbative regime of the heterotic string. 

The anomaly coefficients for the model dependent axions $b_i$ are again calculable for a given compactification. Using the modified Bianchi identity in Eq.~\ref{Biachni-Id-gen_Appendix} one can show that
\begin{equation}
    n_i^{1,2} = \int_X \frac{1}{16\pi^2}\beta_i \wedge \left ( \tr_{1,2} F^2 -\frac{1}{2}\tr R^2 +\frac{1}{3}\sum_r N_r [\Gamma_r]\right ) \,.
\end{equation}

We find that the shape of the linear combinations $\theta_{1,2}$ is slightly modified with respect to the perturbative case with $N_r=0$. However, we conclude that if the 10d gauge group is the usual in weakly coupled heterotic string, $E_8\times E_8$ or $SO(32)$, the presence of NS5-branes and the new (non-perturbative) axions arising from the self-dual 2-form fields $\tilde B_2^{(r)}$ that lives in the worldvolume of NS5-branes do not alter the results derived in the main text. It remains an open question, however, to study axion couplings in the case in which part of the SM gauge group is embedded into the gauge sector non-perturbatively realised by the 5-branes. This is best performed by switching to a dual frame, such as F-theory.

\bibliographystyle{JHEP}

\providecommand{\href}[2]{#2}\begingroup\raggedright\endgroup

\end{document}